\PassOptionsToPackage{cleveref}{jmlrutils}
\PassOptionsToPackage{nameinlink}{cleveref}

\documentclass[pmlr,11pt]{jmlr}
\jmlrproceedings{}{Preprint}
\jmlrvolume{}
\jmlryear{}
\jmlrworkshop{}
\jmlrpages{}

\usepackage{times}
\usepackage{booktabs}
\usepackage{cleveref}

\title[Learning Finite-Memory Binary Sources]{Learning Finite-Memory Binary Sources}

\author{%
 \Name{Maciej Sk\'orski} \Email{maciej.skorski@gmail.com}\\
 \addr University of Luxembourg
}

\usepackage[final]{fixme}
\fxsetup{theme=color, mode=multiuser, layout=inline}
\FXRegisterAuthor{ms}{aMS}{Maciej}

\newcommand{\bits}{\{0,1\}}
\newcommand{\Prob}{\mathbf{P}}
\newcommand{\Exp}{\mathbf{E}}
\newcommand{\KL}{\mathrm{KL}}
\newcommand{\dTV}{d_{\mathrm{TV}}}
\newcommand{\tmix}{t^{\mathrm{mix}}}
\newcommand{\Bern}{\mathrm{Bern}}
\newcommand{\Bin}{\mathrm{Bin}}
\newcommand{\pmin}{\pi_{\min}}
\newcommand{\Cclass}{\mathcal{C}_{\epsilon,d}}
\newcommand{\mgam}{m_{\gamma}}
\newcommand{\dob}{\delta}
\DeclareMathOperator{\osc}{osc}

\usepackage{xcolor}
\usepackage[most]{tcolorbox}
\definecolor{keyblue}{RGB}{70,110,160}
\newcommand{\keyresult}[1]{%
  \begin{tcolorbox}[
    enhanced,
    breakable=false,
    width=\linewidth,
    left=5pt,right=5pt,top=2pt,bottom=2pt,
    boxrule=0pt,
    frame hidden,
    colback=keyblue!3,
    sharp corners,
    before skip=4pt,after skip=4pt
  ]%
  #1
  \end{tcolorbox}%
}

\newcommand{\ub}[2]{\underbrace{#1\vphantom{\sum_{s\geqslant L}\max_{0\leqslant k\leqslant L}}}_{\text{#2}}}

\crefname{appendix}{Appendix}{Appendices}
\crefname{lemma}{Lemma}{Lemmas}
\crefname{proposition}{Proposition}{Propositions}
\crefname{corollary}{Corollary}{Corollaries}
\crefname{remark}{Remark}{Remarks}
\crefname{definition}{Definition}{Definitions}
\Crefname{appendix}{Appendix}{Appendices}
\crefname{subsubsubappendix}{Appendix}{Appendices}
\Crefname{subsubsubappendix}{Appendix}{Appendices}
\crefname{subsubappendix}{Appendix}{Appendices}
\Crefname{subsubappendix}{Appendix}{Appendices}
\begin{document}

\maketitle

\begin{abstract}%
Estimating biased finite-memory sources from a single trajectory is a fundamental problem with practical impact, for example in estimating entropy for cryptography.
We study order-$d$ binary sources whose conditional probabilities have bias at most $\epsilon$, and establish 
\emph{sharp} sample complexity bounds: relative accuracy $\gamma$ with confidence $1-\delta$ needs $\gamma^{-2}\,(\tfrac12-\epsilon)^{-d}\,(\log d+\log\tfrac1\delta)$ samples
up to constants depending on $\epsilon$, with matching upper and lower bounds and no dependence on mixing time or spectral gap.
The proof rests on a sharp bound on expected hitting times of the sliding-window (de Bruijn) chain, which stay of order $1/\pi(w)$ uniformly in $d$ exactly when $\epsilon<\tfrac3{10}$, and on majorization of the window masses by those of independent biased coins.
Applied to published entropy-source validations, the bounds show that their trace lengths may be underpowered: $10^6$ bits cannot resolve all transitions of an $8$-bit model to better than about $4.5\%$ relative error.
\end{abstract}



\section{Introduction}

Learning binary sources with finite memory is of fundamental importance for applications such as cryptographic randomness assessment, lossless compression and sequential prediction
\citep{nist90b,willems1995context,hanjanawu2023}.

Unlike in the equally classical and well-understood \emph{memoryless} model of independent coins, 
here the next-bit probabilities depend on a longer context, and the contexts are visited with exponentially different frequencies, resulting in very diverse reliability for each transition.
How many samples are needed to estimate \emph{reliably and accurately all} of them? 

In this work we settle this problem for sources of bounded bias, pointing out that classical techniques, such as those based on mixing times or spectral gaps, do not yield optimal bounds \Cref{sec:related}.
Our bounds, optimal in the minimax sense, show that the sample size is inversely proportional to the mass of the \emph{rarest} window, inversely quadratic in the desired accuracy, logarithmic in the confidence, and, most importantly, only \emph{double-logarithmic} in the parameters to learn (e.g., in the number of transitions).
More precisely, under next-bit bias at most $\epsilon$, relative accuracy $\gamma$ and confidence $1-\delta$, the sample size splits into three interpretable factors:
\begin{align*}
  n^{\star}
  \;\asymp_\epsilon\;
  \underbrace{\Bigl(\tfrac12-\epsilon\Bigr)^{-d}}_{\text{waiting for the rarest window}}
  \cdot
  \underbrace{\gamma^{-2}}_{\text{accuracy (Fisher information)}}
  \cdot
  \underbrace{\bigl(\log d+\log\tfrac1\delta\bigr)}_{\text{confidence for multiple windows}},
\end{align*}
under the mild restriction $\epsilon<0.3$.

\subsection{Contributions}

Our strategy is to bound the expected hitting times of the sliding-window chain, which controls the visits to a single window, and then to majorize the window masses by those of independent coins, which controls the visits to all windows at once.
We outline the proof plan; key results are highlighted with main ingredients discussed in the main body, the detailed proofs, symbolic calculations or numerical simulations are in the supplementary appendix, which together with the code is also available at \url{https://github.com/maciejskorski/learning-binary-sources.git}. 

\begin{enumerate}
\item \textbf{Sharp overlap of Bernoulli sequences} (\Cref{thm:gap,thm:contraction}).
  Distances and divergences between product distributions are well-studied \citep{talagrand1995missing,smirnov2026tv}, unlike
  the \emph{far-apart} regime: the \emph{overlap} $c_d=1-\dTV$ is what matters for us;
  the larger the overlap, the sooner two starts can be coupled, and the shorter waiting, hitting and coverage times, hence (intuitively) more efficient estimation of transitions.
  For $d$ tosses of coins with biases $\tfrac12\pm\epsilon$, letting $\beta = \sqrt{1-4\epsilon^2}$, $m=\lceil d/2 \rceil$, we prove
  \keyresult{
   $$c_d(\epsilon)\triangleq 1-\dTV\bigl(\Bern(1/2+\epsilon)^{\otimes d},\Bern(1/2-\epsilon)^{\otimes d}\bigr) \asymp \frac{\beta^{2m}}{\sqrt{1+4m\epsilon^{2}}}.$$
  }
  with \textbf{sharp} constants of $1/\sqrt\pi$ from below and $1$ from above. 
  Critically, independent coins are shown to be the extreme overlap case for chain transition laws as well:
  \keyresult{$$\inf_{P\in\Cclass}\bigl(1-\dob(P^d)\bigr)\;=\;c_d(\epsilon).$$} 

  \mswarning{maybe state in dTV and commnent it's Doubling const}

  \item \textbf{Occupancy and hitting times from the oscillation} (\Cref{thm:occ,cor:block}).
  A hitting time is an occupancy in disguise: by the strong Markov property, the visit counter with start at $u$ is behind a start at $w$ by the steps lost while waiting for the first hit.
  Thus it is controlled by the expected visit counts $G_n(\cdot,w)=\sum_{s<n}P^s(\cdot,w)$ of two starts, more precisely by their \emph{oscillation} $\osc G_n(\cdot,w)$, the largest difference between starts.
  We bound this oscillation by a careful split of the series at a block length $L$: the head part keeps the cancellation between the visits of different starts, and the tail is estimated by contraction, because the \emph{column oscillation} of forward transition kernels \emph{decays} geometrically. Specifically, for all $n$,
  \keyresult{\small
   $$\osc G_n(\cdot,w) \leqslant \ub{\max_{k \leqslant L} \osc G_k(\cdot,w)}{head: cancellation} + \ub{\sum_{s\geqslant L}\osc P^s(\cdot,w)}{tail: contraction} \leqslant \max_{k \leqslant L} \osc G_k(\cdot,w) + \frac{L\osc P^{L}(\cdot,w)}{1-\dob(P^L)},$$
  }
  and the same upper bound carries over to the expected hitting times through the fundamental matrix identity.
  Namely, $\pi(w)\Exp_u[\tau_w]=\lim_n\bigl(G_n(w,w)-G_n(u,w)\bigr)\leqslant\sup_n\osc G_n(\cdot,w)$, where $\pi$ is the stationary distribution and $\tau_w$ the hitting time.
  Applications require different regimes of $L$: for shift chains we will use $L=d$, and for expanders $L=\lceil\ln m/\ln(1/\lambda_*)\rceil$.
  
  \msnote{good to go}
  \item \textbf{Hitting without mixing: a sharp phase transition}
  (\Cref{thm:hit,thm:threshold}; \Cref{app:framework}).
  For \emph{any} chain with positive block overlap, expected hitting times are controlled by local quantities, not by mixing, which may be exponentially slow.
  For biased binary sources the worst normalised expected hitting time
  $\mathsf{Hit}_d(\epsilon):=\sup_{P\in\Cclass}\max_{u,w}\pi(w)\,\Exp_u[\tau_w]$ is bounded uniformly
  in the memory $d$ exactly below $\frac3{10}$, and it blows up as the threshold is approached.
  With the distance to the threshold $\Delta:=\bigl|\epsilon-\tfrac3{10}\bigr|$ and
  $\mathsf{Hit}(\epsilon):=\sup_{d}\mathsf{Hit}_d(\epsilon)$, 
  \keyresult{
    $$\begin{cases}
      \Delta^{-1/2}\;\lesssim\;\mathsf{Hit}(\epsilon)\;\lesssim\;\Delta^{-3/2}, & \epsilon<\tfrac3{10},\\[4pt] \mathsf{Hit}_d(\epsilon)\;\gtrsim\;\sqrt d\;e^{3\Delta d}\xrightarrow[d\to\infty]{}\infty\ \ (d\text{ odd}), & \epsilon\geqslant\tfrac3{10}.
    \end{cases}
    $$
  }
  The threshold is caused by a race between the exponential scales of the recurrence time $a^{-d}$ and the crossing time of $\beta^{-d}$ in a bistable chain, which balances at $\epsilon=3/10$.
  \mswarning{working on this, hitting time notation here}
\item \textbf{Final result: sample complexity with $\log d$, not $d$} (\Cref{thm:instance,thm:logd}; \Cref{sec:proof-visits,sec:proof-learning}).
  From an arbitrary initial state, with explicit constants, in \emph{relative} accuracy $\gamma$ (what a min-entropy claim needs,
  at the price $(\tfrac12-\epsilon)^{-1}$ over additive bounds), the failure probability at a window of mass $\pi(w)$
  is at most $\exp(-\kappa_\epsilon\,n\pi(w)\gamma^{2})$ for a constant $\kappa_\epsilon>0$, with no mixing or spectral quantity.
  To handle the visits to \emph{all} windows simultaneously, we majorize the window masses by those of independent biased coins (\Cref{thm:profile}):
  only polynomially many windows are near the minimal mass $b^d$,
  so the union over $2^d$ windows costs $\log d$, not $d$.
  A matching lower bound (Fano's inequality over the $d$ windows with a single $1$, and a two-point argument for the confidence)
  makes the worst case sharp for fixed $0<\epsilon<\tfrac3{10}$:
  \keyresult{
    $$n^{\star}(\gamma,\delta)\;\asymp_\epsilon\;\frac{\log(d+1)+\log\frac1\delta}{\gamma^{2}\,(\frac12-\epsilon)^{d}}.
    $$
  }
  For $\delta\lesssim1/d$, the typical regime, this is $n^{\star}\asymp_\epsilon(\tfrac12-\epsilon)^{-d}\cdot\gamma^{-2}\cdot\ln\tfrac1\delta$: waiting for the rarest window, accuracy, confidence;
  for larger $\delta$ the confidence factor is $\ln d+\ln\frac1\delta$, and the $\log d$ is the price of covering all windows at once.
  No hitting-time constant enters the lower bound (\Cref{thm:logd}).
  The dependence on $\epsilon$ is not settled: the upper bound carries $C(\epsilon)$, which blows up at $\frac3{10}$,
  and we do not show that this is necessary for learning.
  \mswarning{working on that. higlight twice-log of parameter count!}
  \mswarning{praise: the hitting-time cost is additive (\Cref{thm:visits}) and decoupled from the state space, which enters only through $\ln d$.}

\item \textbf{Fast mixing is misleading} (\Cref{prop:extremal}).
  The memoryless chain has the minimum possible mixing time and the minimum possible $\pmin$:
  the minimax lower bound is proved on perturbations of it, so fast mixing alone does not make learning cheap.
  In the worst case the cost is exponential in the memory, with a bias-dependent rate:
  \keyresult{
    $\tmix=d,\qquad\pmin=(\tfrac12-\epsilon)^{d},\qquad \lim_{d\to\infty}\frac{\log_2 n^{\star}}{d}\;=\;1+\log_2\tfrac1{1-2\epsilon}.
    $
    }
\item \textbf{Applications: cryptographic security and expanders} (\Cref{subsec:application}; \Cref{app:audit}).
  \emph{Cryptographic security.} We apply our bounds to published entropy assessments obtained through Markov chain estimation.
  Resolving the min-entropy rate to $\eta$ bit per bit needs, in order of magnitude,
  \keyresult{$n\;\gtrsim\;2^{d}\,\eta^{-2},$}
  and our exact numerical bounds, with explicit constants (\Cref{app:audit}), show that some published traces fall short of the necessary length at their targeted accuracy and confidence.
  \emph{Expanders.} The hitting-time mechanism is not specific to shift chains: on expander graphs the same argument recovers the classical fact that a random walk reaches any vertex from any start in expected time linear in the number of vertices (\Cref{cor:expander}).
\end{enumerate}


\section{Related work}\label{sec:related}

\subparagraph*{Kernel estimation.}
The closest non-asymptotic baselines bound the error of the maximum-likelihood
kernel estimator in terms of a global chain constant. \citet{wolfer2019minimax} give minimax rates for the maximal rowwise $\ell_1$ error
in terms of the pseudo-spectral gap and the minimum stationary probability, and \citet{wolfer2021kernel} a
matrix-level total-variation guarantee with sample size proportional to the mixing time multiplied by a factor depending on stationary transition frequencies, for fixed accuracy and under geometric ergodicity;
\citet{huangli2024} give entrywise Chernoff-type concentration bounds with the iterated Poincaré gap in the exponent, valid for any irreducible chain but with an additional factor that degrades from a poor initial state. \citet{han2018entropy} give non-asymptotic guarantees for entropy-rate estimation of
Markov chains with a large state space, in a model parametrised by a spectral gap,
and \citet{kamath2016entropy} for the min-entropy rate of reversible chains in terms of the relaxation time and the minimum stationary mass.
\citet{hanjanawu2023} study the prediction of Markov chains, including higher-order ones, in expected Kullback--Leibler (KL) risk, with and without a spectral gap.
The kernel-estimation bounds are additive, none is accompanied by a matching lower
bound in our setting, and all treat the $2^d$ windows as generic states.

\subparagraph*{Standard tools and their limits.}
The problem is elementary to state, yet standard tools do not directly yield sharp bounds from bounded bias alone: 
the missing ingredients are (a) quantitative control of visits to each context and (b) a way to combine those guarantees without uniformly treating every context by a union-bound.
The transition probabilities are bounded in $[b,a]$; what varies exponentially are the context frequencies $\pi(w)\in[b^d,a^d]$, products over $d$ steps.

\emph{Spectral theory and perturbation.}
Relaxation-time bounds, such as $\pi(w)\Exp_\pi[\tau_w]\lesssim t_{\mathrm{rel}}$, need a reversible chain with a spectral gap~\citep{aldousfill,levinMarkovChainsMixing2017}.
Sliding-window chains are not reversible, and for such chains eigenvalues can hide slow transients.
The usual fix by symmetrising with $P^{*}P$, does not help either: for the fair source the chain forgets its start after exactly $d$ steps, yet $P^{*}P$ only ever changes the first bit of the window, so it has no gap at all.
Perturbation from the fair source fails too: each transition is within $\epsilon$ of $\frac12$, but this compounds over $d$ steps, since $2^{d}\pi(w)$ ranges from $(1-2\epsilon)^d$ to $(1+2\epsilon)^d$, so no bound uniform in $d$ follows.

\emph{Mixing time.}
Bounds of the form $\Exp_u[\tau_w]\lesssim\tmix/\pi(w)$ are valid but costly here: $\tmix$ can be exponentially large in $d$ (the majority chain of \Cref{thm:threshold}), and our hard instances are small changes of a memoryless source, which mixes in exactly $d$ steps (\Cref{prop:extremal}).

\emph{Hitting times and coverage.}
The naive route waits for $k$ visits at every window:
minorisation $P^d\geqslant b^{d}$ bounds the expected hitting time by $d\,b^{-d}$,
regeneration and Chernoff give accuracy $\gamma$ from $k\asymp\gamma^{-2}\ln(2^{d}/\delta)$ visits,
and a union bound over the $2^d$ windows gives
$n\gtrsim d\,(d+\ln\tfrac1\delta)\,\gamma^{-2}b^{-d}$,
larger than our $\gamma^{-2}b^{-d}(\ln d+\ln\tfrac1\delta)$ by a factor between $d$ and $d^2/\ln d$.
It lower-bounds all windows by the smallest mass $b^d$, the difficulty we handle differently.

\emph{Our route.}
We obtain the missing ingredients through block overlap and occupancy oscillations, which bound the normalised hitting times (\Cref{thm:hit,thm:threshold}), followed by majorization of the stationary window masses by independent coins (\Cref{thm:profile}).

\section{Preliminaries}

\subsection{Notation}

All logarithms are natural unless marked $\log_2$.
The symbols $\lesssim$, $\gtrsim$ and $\asymp$ hide universal positive constants; a subscript indicates permitted dependence.

We follow \citet{levinMarkovChainsMixing2017}.
A Markov chain $(W_t)_{t\geqslant0}$ with transition matrix $P$ lives on a countable state space.
We write $\Prob_x$ and $\Exp_x$ when it starts at $W_0=x$, so that $P^t(x,y)=\Prob_x\{W_t=y\}$.
Distributions are row vectors and functions are column vectors, so that
$\mu P(y)=\sum_x\mu(x)P(x,y)$ and $Pf(x)=\sum_yP(x,y)f(y)=\Exp_xf(W_1)$.
A distribution $\pi$ is \emph{stationary} if $\pi P=\pi$. The \emph{hitting time} of a state $w$ is the random time $\tau_w=\min\{t\geqslant0:\ W_t=w\}$, which is $0$ if the chain starts at $w$,
and the \emph{return time} is $\tau_w^+=\min\{t\geqslant1:\ W_t=w\}$.
Bounds on hitting times are bounds on their expectations $\Exp_u[\tau_w]$ (the \emph{expected hitting time} from $u$), unless stated otherwise.
Let $\tau_w^{(1)}<\tau_w^{(2)}<\cdots$ be the successive visits to $w$, and $N_w(n)=\sum_{t=0}^{n-1}\mathbf 1\{W_t=w\}$ the number of visits among the first $n$ steps (each followed by one new bit).
By the \emph{strong Markov property}, after each visit to $w$ the chain starts afresh from $w$, independently of the past.
The \emph{total variation distance} is $\dTV(\mu,\nu)=\tfrac12\sum_z|\mu(z)-\nu(z)|$,
and the \emph{oscillation} of a real function is $\osc f=\sup f-\inf f$.
The \emph{mixing time} is $\tmix=\min\{t:\ \max_x\dTV(P^t(x,\cdot),\pi)\leqslant\frac14\}$; it appears only for comparison.
The \emph{Dobrushin coefficient}~\citep{dobrushin1956} of $P$ is $\dob(P)=\max_{x,y}\dTV\bigl(P(x,\cdot),P(y,\cdot)\bigr)$,
so $1-\dob(P)=\min_{x,y}\sum_z\min\{P(x,z),P(y,z)\}$ is the \emph{overlap} of two one-step laws with different starts;
it is the contraction factor for oscillation, $\osc(Pg)\leqslant\dob(P)\,\osc g$.
The \emph{expected visit counts} $G_n(x,y)=\sum_{t=0}^{n-1}P^t(x,y)$ give the expected number of visits to $y$
during times $0,\dots,n-1$ from $x$; in equilibrium they equal $n\pi(y)$.
\Cref{app:framework} develops these in detail.

\subsection{The learning problem}\label{subsec:problem}

An order-$d$ source emits each new bit given the context $w\in\bits^d$ of its last $d$ bits, sampling one with probability $p(1\mid w)$.
The class $\Cclass$ consists of the sources of \emph{bias} at most $\epsilon<\tfrac12$, that is,
\begin{equation}\label{eq:bias}
  \bigl|p(1\mid w)-\tfrac12\bigr|\leqslant\epsilon\qquad\text{for all } w,
\end{equation}
and we write $a=\tfrac12+\epsilon$, $b=\tfrac12-\epsilon$, so that $p(v\mid w)\in[b,a]$.
An \emph{estimator} $\hat p$ observes a trajectory of $n$ bits, after an arbitrary initial window, and approximates conditional probabilities $p(\cdot \mid w)$.
For a targeted relative accuracy $\gamma$ and confidence $1-\delta$, we require that, with probability at least $1-\delta$,
\begin{equation}\label{eq:goal}
  \hat p(v\mid w)=(1\pm\gamma)\,p(v\mid w)
  \qquad\text{for all } w\in\bits^d,\ v\in\bits.
\end{equation}
The minimax sample size $n^{\star}(\gamma,\delta)$ is the least $n$ for which some estimator achieves this on every source of the class, from every initial window.
\subsection{The sliding-window chain}\label{subsec:chain}

Let $X=(X_t)_{t\geqslant1}$ be a binary process of order $d$ from the class $\Cclass$ of \Cref{eq:bias},
so that $p(v\mid w)=\Prob\{X_{t+1}=v\mid W_t=w\}\in[b,a]$ for the \emph{window} $W_t=(X_{t-d+1},\ldots,X_t)\in\bits^d$ of the last $d$ bits.
Thus, a source with memory $d$ is a binary chain sliding over $d$-bit windows: in each step the last bit is dropped and a new one added.
This is a Markov chain on $\bits^d$ with the \emph{shift kernel}
\begin{align*}
  P(u,w)\;=\;\begin{cases}p(w_d\mid u), & w=(u_2,\ldots,u_d,w_d),\\ 0, & \text{otherwise,}\end{cases}
\end{align*}
and its transitions are represented by the de Bruijn graph.

The results below follow from the observation that a $d$-step transition of the shift kernel factorizes over a unique path (for $k>d$ steps it is a sum over $2^{k-d}$ paths): starting from $u$, the window after $d$ steps is exactly the string of $d$ new bits, so
$
  P^d(u,w)\;=\;\prod_{j=1}^{d}p\bigl(w_j\mid u_j\cdots u_d\,w_1\cdots w_{j-1}\bigr).
$

\begin{proposition}[Window law bounds]\label{prop:window}
For every $u,w\in\bits^d$,
\[
  b^{d}\;\leqslant\;P^d(u,w)\;\leqslant\;a^{d},
  \qquad\text{and consequently}\qquad
  b^{d}\;\leqslant\;\pi(w)\;\leqslant\;a^{d}.
\]
\end{proposition}

By \Cref{prop:window}, $P^d$ has positive entries, so $P$ is irreducible
and aperiodic with a unique stationary law $\pi$; write $\pmin=\min_w\pi(w)$. We also write $\beta=2\sqrt{ab}=\sqrt{1-4\epsilon^{2}}$.

\mswarning{Kac was in Preliminaries, min-entropy is Verde?}
\begin{proposition}[Kac]\label{prop:kac}
$\Exp_w\tau_w^{+}=1/\pi(w)$ for every $w$.
In particular, all mean return times lie in $[a^{-d},b^{-d}]$, and the shortest one, $1/\max_w\pi(w)$,
is $2$ to the min-entropy of $\pi$.
\end{proposition}

Proofs of \Cref{prop:window,prop:kac} are in \Cref{sec:proof-prelim}.


\section{Results}

Proofs are in \Cref{sec:proofs}; extensions, with their own proofs, are in \Cref{app:extra}; the code and numerical experiments are in the supplementary repository (link in the introduction).

\subsection{Overlap of Bernoulli and Markovian coins}\label{sec:Bernoulli_overlaps}

Classical results control the total variation distance between independent coin product laws, through powers of the Hellinger affinity (Bhattacharyya coefficient) $\beta=\sqrt{1-4\epsilon^{2}}$;
refinements include polynomial prefactors in tails and universal-factor estimates of total variation \citep{talagrand1995missing,smirnov2026tv}.
We focus on the regime where the coins are distinguishable and $\dTV\to1$.
There the \emph{overlap} $c_d=1-\dTV$ decides, since estimates of $\dTV$ are not helpful when it is exponentially small.
\Cref{thm:gap} determines $c_d$ up to \emph{optimal constants}, and \Cref{thm:contraction} shows that independent coins are the extreme (worst) case for Markov dependencies.

\begin{theorem}[Overlap of Bernoulli products]\label{thm:gap}
Let 
\begin{align*}
c_d(\epsilon)\triangleq 1-\dTV\bigl(\Bern(a)^{\otimes d},\Bern(b)^{\otimes d}\bigr) = \sum_{k=0}^{d}\binom dk\min\{a^{k}b^{d-k},b^{k}a^{d-k}\}
\end{align*}
be the distribution overlap of Bernoulli products. For every $d\geqslant1$ and $0\leqslant\epsilon\leqslant\frac12$, with $\beta=\sqrt{1-4\epsilon^{2}}$ and $m=\lceil d/2\rceil$:
\begin{enumerate}
\item[(i)] (parity) $c_{2m-1}(\epsilon)=c_{2m}(\epsilon)$.
\item[(ii)] (exact formula) with $r=2\epsilon$,
  $c_{2m}(\epsilon)=2\Prob\{\Bin(2m-1,b)\geqslant m\}=\displaystyle\frac{\int_r^1(1-u^{2})^{m-1}du}{\int_0^1(1-u^{2})^{m-1}du}$.
\item[(iii)] (sharp bounds)
\begin{align*}
  \frac1{\sqrt\pi}\;\frac{\beta^{2m}}{\sqrt{1+4m\epsilon^{2}}}\;\leqslant\;c_d(\epsilon)\;\leqslant\;\frac{\beta^{2m}}{\sqrt{1+4m\epsilon^{2}}}.
\end{align*}
\end{enumerate}
Both constants are best possible: $1$ is attained as $\epsilon\to0$ and $1/\sqrt\pi\approx0.564$ as $m\to\infty$.
\end{theorem}
\emph{Intuition.} The exponent $\beta^{d}$ is the affinity of the two coins.
If $\epsilon\sqrt d\lesssim1$ the coins are hard to tell apart and this is the whole answer;
if $\epsilon\sqrt d\gg1$ there is an extra factor $1/(\epsilon\sqrt d)$, the chance that a random walk with a small drift ends within $1/\epsilon$ of its centre.
The proof appears in \Cref{sec:proof-overlap}.

\paragraph{From coins to chains.}
We now move from independent coins to dependencies: the same overlap bound survives them, sharply.
\begin{theorem}[Sharp contraction of biased binary chains]\label{thm:contraction}
Consider a sliding-window chain $P\in\Cclass$ of order $d$. Then
\begin{align*}
  \dob(P^d)\;\leqslant\;\dTV\bigl(\Bern(a)^{\otimes d},\Bern(b)^{\otimes d}\bigr)\;=\;1-c_d(\epsilon),
\end{align*}
with $c_d(\epsilon)$ being the overlap of \Cref{thm:gap}.
Equality holds for the chain $p(1\mid w)=b+(a-b)\,w_1$ (where $w_1\in\{0,1\}$
is the oldest bit),
so $c_d(\epsilon)=\inf_{P\in\Cclass}\bigl(1-\dob(P^d)\bigr)$ and the bound cannot be improved.
\end{theorem}
\emph{Intuition.} Two starts are hardest to bring together when one always sits at the top of the allowed range, $p=a$, and the other at the bottom, $p=b$:
their $d$ new bits are then two coins of bias $a$ and $b$.
The chain that copies its oldest bit does exactly this, since for $d$ steps the oldest bit is still the one of the start.
Concavity implies that no other configuration can be worse (worst conditional probabilities at endpoints); the argument is deferred to \Cref{sec:proof-overlap}.

\subsection{Oscillation, occupancy and hitting times}\label{subsec:occ}
We bound occupancy and hitting times of $w$ through the column spread $\osc P^s(\cdot,w)$: differences of visit counts control hitting (\Cref{thm:occ}), and a head--tail split controls the differences (\Cref{cor:block}).
\msnote{good one}
\begin{theorem}[Occupancy and hitting times]\label{thm:occ}
Let $\pi$ be stationary and let $w$ be a state with $\pi(w)>0$
whose \emph{column spread} $\osc P^s(\cdot,w)$ (how much the chance of standing on $w$ at time $s$ still depends on the start) tends to $0$ as $s\to\infty$.
Let $G_n(u,w)$ be the expected visit counts of $w$ from the start $u$.
Then for all states $u$ and all $n\geqslant0$
\begin{align}
  \bigl|G_n(u,w)-n\pi(w)\bigr|&\leqslant\osc G_n(\cdot,w), \label{eq:occ}\\
  \pi(w)\,\Exp_u[\tau_w]&=\lim_{n\to\infty}\bigl(G_n(w,w)-G_n(u,w)\bigr)\leqslant\liminf_{n\to\infty}\osc G_n(\cdot,w). \label{eq:osc-hit}
\end{align}
\end{theorem}

\begin{corollary}[Head and tail]\label{cor:block}
If $\dob(P^L)<1$ for some $L\geqslant1$, then $P$ has a unique stationary law, the column spread contracts, $\osc P^{s+L}(\cdot,w)\leqslant\dob(P^L)\,\osc P^s(\cdot,w)$ for all $s$, so \Cref{thm:occ} applies to every state $w$ with $\pi(w)>0$, and
\begin{align*}
  \sup_{n\geqslant0}\osc G_n(\cdot,w)\;\leqslant\;
  \ub{\max_{0\leqslant k\leqslant L}\osc G_k(\cdot,w)}{head $H_L(w)$}
  +\ub{\sum_{s\geqslant L}\osc P^s(\cdot,w)}{tail}
  \;=:\;B_L(w).
\end{align*}
Here $H_L(w)\leqslant\sum_{s<L}\osc P^s(\cdot,w)$ by subadditivity, and the tail is at most $\sum_{j<L}\osc P^{L+j}(\cdot,w)/(1-\dob(P^L))\leqslant L\,\osc P^L(\cdot,w)/(1-\dob(P^L))$, as the column spreads are nonincreasing.
Hence \emph{both} the occupancy at every finite $n$ and the expected hitting time are bounded by $B_L(w)$:
$|G_n(u,w)-n\pi(w)|\leqslant B_L(w)$ and $\pi(w)\Exp_u[\tau_w]\leqslant B_L(w)$.
\end{corollary}

\emph{Intuition.} A walker started at $u$ is behind one started at $w$ by the steps lost before the first hit, so hitting times are differences of visit counts, bounded by sums of column spreads.
The first $L$ (the head) are kept for cancellations; the rest (the tail) contract geometrically.
Proofs: \Cref{sec:proof-occ}.
\mswarning{simplify intuition}

\subsection{Hitting times and occupancy of sliding-window chains}

We now combine our sharp results on the overlap of Bernoulli sequences (\Cref{sec:Bernoulli_overlaps}),
together with the oscillation control of occupancy and hitting times (\Cref{subsec:occ}),
to obtain accurate bounds on the expected hitting times of biased shift chains:
below bias $\frac3{10}$, the normalised hitting times $\pi(w)\Exp_u[\tau_w]$ stay bounded independently of the memory $d$.

\begin{theorem}[Hitting times and occupancy of biased binary sources]\label{thm:hit}
  Consider a binary chain of order $d$ in $\Cclass$, i.e.\ with $p(1\mid s)\in[b,a]$ for all windows $s$.
  Let $c_d(\epsilon)$ be the overlap of \Cref{thm:gap}.
Put
\begin{align*}
B_d(\epsilon) \triangleq \frac{1-a^{d}}{1-a}+\frac{d\,(a^{d}-b^{d})}{c_d(\epsilon)}.
\end{align*}
Then for all $u$, $w$ and $n$, with the visit counts $G_n$ of \Cref{thm:occ},
\begin{align*}
  \osc G_n(\cdot,w) &\leqslant B_d(\epsilon),
  &
  \pi(w)\,\Exp_u[\tau_w] &\leqslant B_d(\epsilon).
\end{align*}
For $\epsilon<\frac3{10}$ this is \emph{uniform in $d$}:
$B_d(\epsilon)\leqslant C(\epsilon):=\sup_{d\geqslant1}B_d(\epsilon)<\infty$,
where $C(\epsilon)\leqslant\frac25\,(\frac3{10}-\epsilon)^{-3/2}$ (\Cref{sec:proof-hit}; exact values in \Cref{tab:C}).
\end{theorem}
\begin{table}[t]
\centering
\begin{tabular}{l|ccccccc}
$\epsilon\leqslant$ & 0.05 & 0.10 & 0.15 & 0.20 & 0.25 & 0.28 & 0.29\\\hline
$C(\epsilon)\leqslant$ & 2.31 & 2.84 & 3.71 & 5.34 & 10.61 & 30.42 & 76.37
\end{tabular}
\caption{Verified upper bounds on $\sup_{d}B_d(\epsilon)$ (\Cref{sec:proof-hit}). 
}
\label{tab:C}
\end{table}

\emph{Intuition.} The first term ($\leqslant1/b$) is the memory of the first $d$ steps; the second is the cost of merging afterwards, governed by the overlap $c_d(\epsilon)$ of \Cref{thm:contraction}.
It grows like $(a/\beta)^{d}$ up to a polynomial factor in $d$, so $B_d$ is bounded exactly when $a<\beta$, i.e.\ $\epsilon<\frac3{10}$ (the race is explained after \Cref{thm:threshold}).
Proof: \Cref{sec:proof-hit}.
The threshold $\frac3{10}$ is not a proof limitation, but rather a transition in the behaviour of hitting times.


\begin{theorem}[Sharp transition at $\epsilon=0.3$ for normalised hitting times]\label{thm:threshold}
For $0\leqslant\epsilon<\tfrac12$ and $\mathsf{Hit}(\epsilon)=\sup_{d}\sup_{P\in\Cclass}\max_{u,w}\pi(w)\Exp_u[\tau_w]$,
\begin{align*}
  \mathsf{Hit}(\epsilon)<\infty\quad\Longleftrightarrow\quad\epsilon<\tfrac3{10}.
\end{align*}
More precisely, for odd $d$ let $M_d\in\Cclass$ be the \emph{majority chain}:
$p(1\mid w)=a$ if $w$ has more ones than zeros, and $p(1\mid w)=b$ otherwise.
Then
\begin{align*}
  \pi(1^{d})\,\Exp_{0^{d}}[\tau_{1^{d}}]\;\geqslant\;\frac{\sqrt{2\pi}\,\epsilon}{4\beta}\,\sqrt d\,\Bigl(\frac a\beta\Bigr)^{d},
\end{align*}
which tends to infinity with $d$ exactly when $a\geqslant\beta$,
i.e.\ $\epsilon\geqslant\tfrac3{10}$ (at $\epsilon=\tfrac3{10}$ one has $a=\beta=0.8$).
\end{theorem}

\emph{Intuition.} The majority chain is \emph{bistable}: its majority rule traps the chain in the well it is in, a well of mostly zeros or of mostly ones.
To reach $1^d$ from $0^d$ it must first \emph{escape} the zeros well, which needs a window with a majority of ones from a coin of bias $b$: probability about $\beta^{d}$ per step, so a waiting time about $\beta^{-d}$.
The target $1^d$ has mass about $a^{d}$, so the normalised time is about $(a/\beta)^{d}$, unbounded exactly when $a\geqslant\beta$.
The proof appears in \Cref{sec:proof-threshold}.



\subsection{Bounds on multiple visits}\label{subsec:visits}

Good estimation of $p(\cdot\mid w)$ requires a number of visits to each window $w$.
The time until every state is visited $k$ times, known as the $k$-cover time~\citep{chan2021cover}, is the right tool here.

\paragraph{Multiple visits to one state.}

We first prove that, during $n$ steps, each window $w$ is visited at least $\gtrsim n\pi(w)$ times (not much less than expected) with overwhelming probability $1-\exp\bigl(-\Omega(n\pi(w))\bigr)$.
For our class the bound holds uniformly over $d$ when $\epsilon<\frac3{10}$ (\Cref{thm:hit}), which gives the Chernoff-type guarantee \eqref{eq:visits}, with explicit constants. In addition, the argument isolates the hitting-time cost as an additive term (a naively used hitting time incurs a multiplicative cost); this avoids a polynomially large prefactor $C(\epsilon)\lesssim(\frac3{10}-\epsilon)^{-3/2}$ in the sample size. 

\begin{theorem}[Repeated visits to one state]\label{thm:visits}
Fix a window $w$ and let $t_w=\max_y\Exp_y[\tau_w]$ be the expected time to hit $w$ from the worst start.
Then, for every $\eta\in(0,1)$, with probability at least $1-\eta$ the $k$-th visit to $w$ happens by time
\begin{align*}
  \tau_w^{(k)}\;\leqslant\;1.3\,\frac{k-1}{\pi(w)}+6.5\,t_w\ln\frac2\eta .
\end{align*}
\end{theorem}

For $\epsilon<\frac3{10}$ we have $t_w\leqslant B_d/\pi(w)$ with $B_d\leqslant C(\epsilon)$ (\Cref{thm:hit}), and the exponential-tail bound in the proof of the theorem gives, for every chain in $\Cclass$, window $w$, start and $n,k\geqslant1$,
\begin{align}
  \Prob\bigl\{N_w(n)<k\bigr\}\;\leqslant\;\frac65\,\exp\Bigl(-\frac{(n-1)\,\pi(w)-1.3\,(k-1)}{6.5\,B_d}\Bigr).\label{eq:visits}
\end{align}
So, with probability at least $1-\frac65\exp\bigl(-\frac{(n-1)\pi(w)}{13\,B_d}\bigr)$ the window is visited at least $\frac{(n-1)\pi(w)}{2.6}$ times.

\emph{Intuition.}
After the first visit the chain is at $w$, so each revisiting costs one return time, of mean $1/\pi(w)$ (the first term; the constant $1.3$ can be further improved towards $1$).
The delay before the first visit, of mean at most $t_w$, contributes only once (the second term).
Thus, the costs add rather than multiply.
The proof is in \Cref{sec:proof-visits}.

\paragraph{Multiple visits to \emph{all} states.}
Accurate estimation needs multiple visits to all windows at once; since their masses are unknown and can vary a lot across windows $w$,
the bounds are not directly summable. Indeed, a uniform bound, which treats every window with the smallest mass $b^d$, would overshoot the true failure probability by a factor exponential in $d$,
resulting in a sample size linear rather than logarithmic in $d$.

We handle this by majorizing, once again, by independent coins: the window masses of every chain of the class are majorized by those of $d$ independent biased coins.
This buys $\log d$ instead of $d$: for independent coins the masses are $b^d(a/b)^k$ with exactly $\binom dk$ windows at level $k$, only polynomially many windows near the smallest mass.
Just as in coupon collection with unequal probabilities, the cost is driven by how many coupons are rare, rather than how many are there.

\begin{theorem}[Majorization by independent coins]\label{thm:profile}
Let $P$ be a chain of order $d$ whose conditional probabilities lie in the box $[b,a]$, $a=1-b$. For every convex function $\varphi$,
\begin{align*}
  \sum_{w}\varphi\bigl(\pi(w)\bigr)\;\leqslant\;\sum_{k=0}^{d}\binom dk\,\varphi\bigl(a^{k}b^{d-k}\bigr),
\end{align*}
with equality for the memoryless chain.
\end{theorem}

\emph{Intuition.} The failure bound is a convex function of the mass (an exponential), and a convex sum is largest when the masses are spread out as much as possible.
We then study the stationary law as a binary tree: at each branch the mass is split between the two next bits with probabilities in $[b,a]$.
By convexity, the worst tree splits every node as $(a,b)$, which is the memoryless chain.
In fact, the box need not be symmetric: if only $p(1\mid w)\in[\beta,\alpha]$, the theorem holds with $a:=\max\{\alpha,1-\beta\}$ and $b:=1-a$.
The proof is in \Cref{sec:proof-visits}.

\begin{corollary}[Summing over windows for our class]\label{cor:union}
Let $P\in\Cclass$ with $\epsilon>0$ ($b=\frac12-\epsilon$, $a=\frac12+\epsilon$), and let $c,t>0$.
Suppose that each window $w$ has an "exceptional" event (for instance, too few visits) of probability at most $c\,e^{-t\pi(w)}$.
If $t\,b^{d}\geqslant\frac b{2\epsilon}(1+\ln d)$, then the probability that some window is exceptional is at most $1.445\,c\,e^{-t\,b^{d}}$.
\end{corollary}

\emph{Intuition.} By \Cref{thm:profile} we may count windows as for independent coins: level $k$ has at most $d^{k}$ windows, but under the stated condition its windows are so much heavier that level $k$ adds at most $e^{-k}$ times the failure probability of the lightest window. 
The sum is essentially geometric, so all windows together cost a constant factor, and the price of the union is the condition $t\,b^{d}\gtrsim\ln d$, which is the $\log d$ of the sample size.

\subsection{Learning bounds}

\paragraph{The estimator.}
Fix a window $w$ and let $\xi_j$ be the bit following the $j$-th visit to $w$. 
The estimator is the standard maximum-likelihood one, the empirical frequency of these bits (\Cref{alg:estimator}). 

\begin{algorithm}[ht]
\caption{Estimator of transition probabilities of a sliding-window chain}\label{alg:estimator}
\KwIn{initial window and $n$ new bits $x_{1-d},\dots,x_n$, window $w$, bit $v$}
\KwOut{estimate $\hat p(v\mid w)$}
$S\gets\{0\leqslant t<n:\ (x_{t-d+1},\dots,x_t)=w\}$ \tcp*{visits to $w$}
$N\gets|S|$\;
\lIf{$N=0$}{\Return $\frac12$}
\Return $\frac1N\sum_{t\in S}\mathbf 1\{x_{t+1}=v\}$\;
\end{algorithm}

Each bit $\xi_j$ is a fresh draw from $p(\cdot\mid w)$, whatever happened before (strong Markov property), so the $\xi_j$ are i.i.d.
The estimator uses the trajectory only, with no knowledge of the bias, the stationary law, or mixing and hitting times.
Combined with the multiple-visit bounds of \Cref{subsec:visits} (\eqref{eq:visits} for one window, \Cref{thm:profile} for all of them), this gives the final result:

\begin{theorem}[Sample complexity, upper bound]\label{thm:instance}
Let $\epsilon<\tfrac12$ and $\gamma\leqslant1$, let $B_d=B_d(\epsilon)$ be the constant of \Cref{thm:hit}, and let $\mgam=\frac{1+2\epsilon}{(1-2\epsilon)\gamma^{2}}\asymp_\epsilon\gamma^{-2}$ be the number of flips needed to learn the rarer bit to relative accuracy $\gamma$.
For every chain in $\Cclass$, every window $w$ and every $n\geqslant1$, the estimator has relative error above $\gamma$ at $w$ with probability at most
\begin{align*}
  4\exp\Bigl(-\frac{n\,\pi(w)}{8\,(\mgam+B_d)}\Bigr).
\end{align*}
Consequently, for $0<\epsilon<\tfrac3{10}$ and $C(\epsilon)=\sup_dB_d(\epsilon)$, \Cref{eq:goal} holds w.p. $\geqslant 1-\delta$ as soon as
\begin{align*}
  n\;\geqslant\;\frac{8\bigl(\mgam+C(\epsilon)\bigr)}{b^{d}}\,\max\Bigl\{\ln\frac6\delta,\ \frac b{2\epsilon}\,(1+\ln d)\Bigr\}.
\end{align*}
\end{theorem}

\emph{Intuition.} A window of mass $\pi(w)$ is seen about $n\pi(w)$ times, each visit is one coin flip, and \Cref{thm:hit} says no start delays the visits by more than a constant factor on the scale $1/\pi(w)$.
So a window is learned once $n\pi(w)\gtrsim\mgam+B_d$, and the union over windows adds only $\log d$ (\Cref{sec:proof-visits}).
The proof uses \Cref{thm:hit} only through \Cref{thm:visits}, so any bound on the hitting time $t_w$ would do.
The proof is in \Cref{sec:proof-learning}. 

It remains to show that these costs cannot be avoided. 
We show a matching lower bound that applies to every estimator. 
The key property used in the proof is that trajectory divergence equals the expected number of visits to the modified row times its per-visit divergence (\Cref{lem:kl}).



\begin{theorem}[Lower bound with the $\log d$ term]\label{thm:logd}
Let $0<\epsilon<\tfrac12$, $d\geqslant1$, $\gamma\leqslant\min\{\epsilon/4,\tfrac1{12}\}$ and $\delta\leqslant\tfrac14$.
Then every procedure that satisfies \Cref{eq:goal} with probability at least $1-\delta$ on every chain of $\Cclass$, from the initial state $1^d$, needs
\begin{align*}
  n\;\geqslant\;\frac{c}{\gamma^{2}\,b^{d}}\Bigl[\ln(d+1)+\ln\frac1\delta\Bigr],\qquad c=\tfrac1{20}.
\end{align*}
Together with \Cref{thm:instance}, for fixed $0<\epsilon<\tfrac3{10}$ and $\gamma,\delta$ as above,
\begin{align*}
  n^{\star}(\gamma,\delta)\;\asymp_\epsilon\;\frac{\log(d+1)+\log\frac1\delta}{\gamma^{2}\,b^{d}}.
\end{align*}
No hitting-time constant $C(\epsilon)$ enters the lower bound, and it holds for every bias $\epsilon<\frac12$:
the threshold $\frac3{10}$ limits only the matching upper bound.
\end{theorem}

\emph{Intuition.} The memoryless chain with $\Prob\{0\}=b$ has $d$ windows with a single $1$, all of mass $\asymp b^d$.
Raising the zero-probability at one of them by the factor $1+2.2\gamma$ gives $d$ alternatives,
each costing about $\gamma^{-2}$ visits to detect; finding \emph{which} one changed needs $\ln d$ times more information (Fano), not $d$ times.
The proof, which also covers \Cref{thm:instance}, can be found in \Cref{sec:proof-learning}.

\subsection{Mixing is misleading; consequence for practice}

The rarest possible window already occurs in a memoryless source, so rapid mixing does not remove the learning cost.

\begin{proposition}[Extremal chain: mixing is misleading]\label{prop:extremal}
Over $\Cclass$, $\pmin$ is minimised by the \emph{memoryless} chain
$p(1\mid w)\equiv\tfrac12+\epsilon$, for which
$\pi(w)=(\tfrac12+\epsilon)^{|w|}(\tfrac12-\epsilon)^{d-|w|}$ and
$\pmin=\pi(0^d)=b^{d}$, meeting the floor of \Cref{prop:window} with equality. This
chain has $\tmix=d$ exactly, the minimum possible over $\Cclass$.
The minimax lower bound of \Cref{thm:logd} is proved on perturbations of this chain,
so the worst-case order of the learning cost already occurs among near-memoryless sources, which mix fast:
fast mixing alone does not imply cheap learning.
\end{proposition}

\emph{Consequence.} For fixed $0<\epsilon<\tfrac3{10}$ and $\gamma,\delta$, \Cref{thm:instance,thm:logd} give $\lim_{d\to\infty}d^{-1}\log_2n^{\star}=1+\log_2\frac1{1-2\epsilon}\approx1+2.885\,\epsilon$: bias costs about $2.9\epsilon$ bits per unit of memory, accuracy only $2\log_2\frac1\gamma$ once.
For realistic trace lengths, full-kernel validation is therefore infeasible beyond $d\approx16$--$20$ (see the audit in \Cref{app:audit}).

\subsection{Applications: cryptographic security and expanders}\label{subsec:application}

\paragraph{Cryptographic security: validating an entropy source.}
A hardware entropy source is certified by fitting a Markov chain of memory $d$ to a captured trace and computing a min-entropy rate from the fitted kernel, as in NIST SP~800-90B~\citep{nist90b} and in published evaluations~\citep{hamburg2012ivy}.
What is missing is the statistics of the fit: how long the trace must be for the \emph{rare} windows, to which a min-entropy bound is most sensitive, to be seen often enough.
The rarest window gets about $n2^{-d}$ observations; applied to published validations (OpenTRNG, a multi-ring-oscillator source, Intel's Ivy Bridge source), our bounds show that traces of $10^{6}$ bits fall short by a factor $23$ at $d=8$ (\Cref{app:audit}).
The bounds hold for sources in $\Cclass$. 

\paragraph{Expanders.}
The hitting-time mechanism is not specific to shift chains. As a fast-mixing example,
let $G$ be a connected non-bipartite $k$-regular graph on $m\geqslant4$ vertices
and $P$ its simple random walk, $\pi$ is uniform.
Let $\lambda_*\in(0,1)$ be the largest absolute value of an eigenvalue of $P$ other than $1$.

\begin{corollary}[Expanders]\label{cor:expander}
With $L=\lceil\ln m/\ln(1/\lambda_*)\rceil$, for all vertices $u,w$,
\begin{align*}
\pi(w)\,\Exp_u[\tau_w]=\frac1m\Exp_u[\tau_w]\leqslant\frac{1}{1-\lambda_*}+\frac{5L}{m} .
\end{align*}
In particular, for a family of expanders, whose spectral gap $1-\lambda_*$ is bounded below,
$\Exp_u[\tau_w]\lesssim m$ uniformly in $u,w$.
\end{corollary}

\emph{Intuition.} By \Cref{thm:occ}, $\pi(w)\Exp_u[\tau_w]$ is a difference of visit counts of two starts. After $L\asymp\log m/(1-\lambda_*)$ steps the walk is close to uniform from every start, so only the $O(L/m)$ visits to $w$ within a block and the returns to $w$ (bounded through the gap) remain.
For a bounded gap this gives $\Exp_u[\tau_w]\lesssim m$, the order of the return time $m=1/\pi(w)$ (Kac) and of the classical bound $t_{\mathrm{rel}}/\pi(w)$.
The proof, an application of the block bound \Cref{thm:overlap}, is in \Cref{app:expander}.

\section*{Conclusions}
Learning an order-$d$ binary source of bias at most $\epsilon<\tfrac3{10}$ to relative accuracy $\gamma$ with confidence $1-\delta$ takes $\gamma^{-2}(\tfrac12-\epsilon)^{-d}(\log d+\log\tfrac1\delta)$ samples, up to constants depending on $\epsilon$; this is optimal in the minimax sense and involves no mixing time or spectral gap.
As for practical consequences, our bounds suggest that trace lengths in published entropy-source validations may be underpowered; for theory, extending the matching upper bound beyond $\epsilon=\tfrac3{10}$, where hitting times are no longer of order $1/\pi(w)$, is the main open question.

\section*{AI Disclosure}
AI assistant (Claude, Anthropic) was used for minor editorial polishing, related-work search, and coding numerical simulations. 
The authors reviewed all AI output, and take full responsibility for the entire article content.

\clearpage

\bibliography{references}

\appendix

\clearpage 

\section{Proofs}
\label[appendix]{sec:proofs}
In the proofs we abbreviate the column spread by $\Omega_s(w):=\osc P^s(\cdot,w)=\max_xP^s(x,w)-\min_xP^s(x,w)$.


The proofs follow the order of the results, and no mixing quantity enters any of them.
After the preliminary window-law and return-time facts, we prove the sharp overlap and contraction of biased chains (\Cref{thm:gap,thm:contraction}),
then, in one subsection, the occupancy and hitting-time bounds from the column oscillation (\Cref{thm:occ,cor:block}),
the uniform hitting times of sliding-window chains (\Cref{thm:hit}, with the numerical constant of \Cref{thm:C}),
and the threshold, proved with the majority chain (\Cref{thm:threshold}).
Then come the repeated-visit and majorization bounds (\Cref{thm:visits,thm:profile}),
and the learning bounds (\Cref{thm:instance,thm:logd}).
The proofs of the additional results (\Cref{app:extra}) are collected next to their statements.

\subsection{Window laws and return times}\label[appendix]{sec:proof-prelim}

\begin{proof}[Proof of \Cref{prop:window}]
Each of the $d$ factors in the product lies in $[b,a]$.
Since $\pi=\pi P^d$, we get $\pi(w)=\sum_u\pi(u)P^d(u,w)$, which lies in the same interval.
\end{proof}

\begin{proof}[Proof of \Cref{prop:kac}]
The first claim is Kac's formula~\citep{kac1947}; the range follows from \Cref{prop:window}.
The shortest mean return time is $1/\max_w\pi(w)=2^{H_\infty(\pi)}$, where $H_\infty(\pi)=-\log_2\max_w\pi(w)$.
\end{proof}

\subsection{Overlap of Bernoulli and Markovian coins}\label[appendix]{sec:proof-overlap}

\begin{proof}[Proof of \Cref{thm:gap}]
Write $\rho=a/b$, so $t=b/a=1/\rho$, and $\beta=2\sqrt{ab}$, so $4ab=\beta^{2}$.
The cases $\epsilon=0$ ($c_d=1$) and $\epsilon=\frac12$ ($c_d=0=\beta$) are trivial,
so let $0<\epsilon<\frac12$.
For $z\in\{0,1\}^{d}$ with $k$ ones put $A(z)=a^{k}b^{d-k}$ and $B(z)=b^{k}a^{d-k}$,
so $c_d=\sum_z\min\{A(z),B(z)\}$.

\emph{(i) Parity.} Let $d$ be odd.
Then $A(z)/B(z)=\rho^{2k-d}$ with $2k-d$ odd,
so either $A/B\geqslant\rho$ or $A/B\leqslant1/\rho$.
Appending a bit multiplies $A$ by $a$ or $b$ and $B$ by $b$ or $a$, hence
\begin{align*}
  \sum_{x\in\{0,1\}}\min\{A(zx),B(zx)\}\;=\;\min\{aA,bB\}+\min\{bA,aB\}.
\end{align*}
If $A/B\geqslant\rho$ both minima are the $B$-terms and the sum is $(a+b)B=B=\min\{A,B\}$;
if $A/B\leqslant1/\rho$ both are the $A$-terms and the sum is $A=\min\{A,B\}$.
Summing over $z$ gives $c_{d+1}=c_d$.

\emph{(ii) Exact formula.} Let $d=2m-1$ and $z\in\{0,1\}^{d}$ with $k$ ones.
If $k\geqslant m$ then $A(z)>B(z)$ and the minimum is $B(z)=b^{k}a^{d-k}$;
the terms with $k<m$ are their mirror images under $k\mapsto d-k$.
Hence $c_{2m-1}=2\Prob\{\Bin(2m-1,b)\geqslant m\}$, which equals $c_{2m}$ by (i).
The binomial tail is a Beta probability (the regularised incomplete beta function, \citealp[\S8.17]{dlmf}):
$\Prob\{\Bin(2m-1,b)\geqslant m\}=\Prob\{\mathrm{Beta}(m,m)\leqslant b\}$.
Let $U=1-2\,\mathrm{Beta}(m,m)$, symmetric on $(-1,1)$ with density proportional to $(1-u^{2})^{m-1}$. Then
\begin{align*}
  c_{2m}\;=\;\Prob\{|U|\geqslant r\}\;=\;N_m\int_r^{1}(1-u^{2})^{m-1}du,\qquad r=2\epsilon,\quad N_m=\frac{(2m-1)!}{4^{m-1}((m-1)!)^{2}}, \tag{$*$}
\end{align*}
where $N_m=\bigl(4^{m-1}\mathrm{B}(m,m)\bigr)^{-1}$ is the Beta normalising constant.

\emph{(iii).} Put $K_m=\binom{2m}m4^{-m}$, $\kappa_m=K_m\sqrt{m+1}$ and, with $r=2\epsilon$,
\begin{align*}
  T(r)\;&=\;N_m\int_r^1(1-u^{2})^{m-1}du\;=\;c_{2m},\\
  R(r)\;&=\;(1-r^{2})^{m}(1+mr^{2})^{-1/2}\;=\;\frac{\beta^{2m}}{\sqrt{1+4m\epsilon^{2}}}.
\end{align*}
We show $\kappa_mR\leqslant T\leqslant R$ on $[0,1]$, and then $1/\sqrt\pi\leqslant\kappa_m<1$.

\emph{Derivatives.} $T(0)=R(0)=1$ by the normalisation in $(*)$, and $T(1)=R(1)=0$. Further,
\begin{align*}
  T'(r)\;&=\;-N_m(1-r^{2})^{m-1},\\
  R'(r)\;&=\;-(1-r^{2})^{m-1}q(r),\qquad q(r)\;=\;\frac{m\,r\,\bigl(3+(2m-1)r^{2}\bigr)}{(1+mr^{2})^{3/2}}.
\end{align*}
The function $q$ is increasing on $(0,1)$, since
\begin{align*}
  \frac{r\,q'(r)}{q(r)}\;=\;\frac{3(1-r^{2})}{\bigl(3+(2m-1)r^{2}\bigr)(1+mr^{2})}\;>\;0,
\end{align*}
with $q(0)=0$ and $q(1)=2m/\sqrt{m+1}$. Since $N_m=2mK_m$, this reads $\kappa_mq(1)=N_m$.

\emph{The constants.} The ratio of consecutive terms is $\kappa_{m+1}/\kappa_m=\frac{2m+1}{2m+2}\sqrt{\frac{m+2}{m+1}}<1$,
so $\kappa_m$ decreases from $\kappa_1=1/\sqrt2$, in particular $\kappa_m<1$; its limit is $1/\sqrt\pi$ by Stirling's formula, hence $\kappa_m\geqslant1/\sqrt\pi$.
Therefore $q(1)>N_m$.

\emph{Lower bound.} $(T-\kappa_mR)'=(1-r^{2})^{m-1}\bigl[\kappa_mq(r)-N_m\bigr]\leqslant0$,
because $\kappa_mq(r)\leqslant\kappa_mq(1)=N_m$.
So $T-\kappa_mR$ decreases to its value $0$ at $r=1$, hence is $\geqslant0$.

\emph{Upper bound.} $(R-T)'=(1-r^{2})^{m-1}\bigl[N_m-q(r)\bigr]$.
As $q$ increases from $0$ to $q(1)>N_m$, this derivative is positive and then negative, once.
So $R-T$ increases and then decreases, with values $0$ at $r=0$ and $r=1$, hence is $\geqslant0$.

\emph{Sharpness.} For fixed $m$, $T/R\to1$ as $\epsilon\to0$.
As $\epsilon\to\frac12$ we have $T/R=\sqrt{1+mr^{2}}\;T/(1-r^{2})^{m}$ and, by l'Hôpital's rule, $T/(1-r^{2})^{m}\to\lim\frac{N_m(1-r^{2})^{m-1}}{2mr(1-r^{2})^{m-1}}=\frac{N_m}{2m}=K_m$, so $T/R\to\kappa_m$.
Thus $1$ is the best upper constant for every $m$, and $\kappa_m\downarrow1/\sqrt\pi$ is the best lower one.
\end{proof}

\begin{proof}[Proof of \Cref{thm:contraction}]
Let $A_n=\Bern(a)^{\otimes n}$, $B_n=\Bern(b)^{\otimes n}$ and, for $s,t\geqslant0$,
\begin{align*}
  F_n(s,t):=\sum_{z\in\{0,1\}^{n}}\min\{s\,A_n(z),\,t\,B_n(z)\}.
\end{align*}
$F_n$ is concave (a sum of minima of linear functions) and positively homogeneous,
$F_n(\lambda s,\lambda t)=\lambda F_n(s,t)$. Splitting according to the first bit,
\begin{align}
  F_{n+1}(s,t)&=F_n(sa,tb)+F_n(sb,ta). \label{eq:F-split}
\end{align}
Moreover $F_{n+1}\leqslant F_n$, because $\min\{u_1,v_1\}+\min\{u_2,v_2\}\leqslant\min\{u_1+u_2,v_1+v_2\}$ and $a+b=1$:
forgetting the last bit can only increase the overlap.
Call a law $\mu$ on $\{0,1\}^n$ \emph{admissible} if all its next-bit conditional probabilities lie in $[b,a]$.

\emph{Claim.} For admissible $\mu,\nu$ on $\{0,1\}^n$ and all $s,t\geqslant0$,
\begin{align}
  \sum_{z}\min\{s\,\mu(z),\,t\,\nu(z)\}\;\geqslant\;F_n(s,t). \label{eq:claim}
\end{align}
For $n=0$ both sides equal $\min\{s,t\}$. For the step $n\to n+1$,
let $p,q\in[b,a]$ be the first-bit probabilities of $\mu,\nu$;
given the first bit, the remaining bits are admissible laws of length $n$.
Conditioning on the first bit and applying the induction hypothesis to the two branches,
which carry the masses $sp,tq$ and $s(1-p),t(1-q)$, gives
\begin{align*}
  \sum_{z}\min\{s\,\mu(z),\,t\,\nu(z)\}\;\geqslant\;\Phi(p,q):=F_n(sp,tq)+F_n\bigl(s(1-p),t(1-q)\bigr).
\end{align*}
Now $\Phi$ is concave on the rectangle $[b,a]^2$, so its minimum is attained at a corner.
At $(a,b)$ and $(b,a)$, \Cref{eq:F-split} gives $\Phi=F_{n+1}(s,t)$. At $(a,a)$ and $(b,b)$,
homogeneity gives $\Phi=(a+b)F_n(s,t)=F_n(s,t)\geqslant F_{n+1}(s,t)$.
Hence $\Phi\geqslant F_{n+1}(s,t)$ on the whole rectangle, which proves \Cref{eq:claim} for $n+1$.

\emph{Contraction.} For windows $x,y$,
the laws $\mu=P^d(x,\cdot)$ and $\nu=P^d(y,\cdot)$ of the next $d$ bits are admissible:
the $j$-th conditional probability is $p(1\mid x_j\cdots x_d\,v_1\cdots v_{j-1})\in[b,a]$.
Applying \Cref{eq:claim} with $s=t=1$ gives $1-\dTV(\mu,\nu)\geqslant F_d(1,1)=1-\dTV(A_d,B_d)=c_d(\epsilon)$ (a string with $k$ ones has $A_d(z)=a^{k}b^{d-k}$ and $B_d(z)=b^{k}a^{d-k}$).
Maximising over $x,y$ gives the bound on $\dob(P^d)$.

\emph{Equality.} For the chain $p(1\mid w)=a$ if $w_1=1$ and $b$ if $w_1=0$,
the window at step $j\leqslant d$ is $x_j\cdots x_d\,v_1\cdots v_{j-1}$,
whose oldest bit is $x_j$. From $x=1^d$ all conditionals equal $a$,
so the next $d$ bits have law $A_d$;
from $y=0^d$ they have law $B_d$. Hence $\dob(P^d)\geqslant\dTV(A_d,B_d)$.
\end{proof}

\subsection{Occupancy, hitting times and the threshold}

\subsubsection{Oscillation and occupancy}\label[appendix]{sec:proof-occ}

\begin{lemma}[Oscillation contraction]\label{lem:osc}
For a transition matrix $Q$ and any bounded function $g$,
$\osc (Qg)\leqslant\dob(Q)\osc (g)$.
In particular, for every state $w$ and $s,L\geqslant0$,
\begin{align}\label{eq:contract}
\Omega_{s+L}(w)\leqslant\dob(P^L)\,\Omega_s(w).
\end{align}
\end{lemma}

\emph{Intuition.} Two walks starting at $x$ and $y$ after $L$ steps forget a
fraction $1-\dob(P^L)$ of where they begin, so the averages of $g$ seen from $x$ and from $y$ differ by at most the remaining
fraction $\dob(P^L)$.

\begin{proof}[Proof of \Cref{lem:osc}]
Let $\mu=Q(x,\cdot)$ and $\nu=Q(y,\cdot)$. Since $\sum_z(\mu(z)-\nu(z))=0$,
we may subtract $\min g$ from $g$:
\begin{align*}
Qg(x)-Qg(y)&=\sum_z\bigl(g(z)-\min g\bigr)\bigl(\mu(z)-\nu(z)\bigr)\\
&\leqslant\osc (g)\sum_z\bigl(\mu(z)-\nu(z)\bigr)_+=\osc (g)\,d_{TV}(\mu,\nu),
\end{align*}
and we maximize over $x,y$. For \Cref{eq:contract} use $P^{s+L}(\cdot,w)=P^L g$ with $g=P^s(\cdot,w)$.
\end{proof}

\begin{proof}[Proof of \Cref{thm:occ}]
\emph{Convergence of the columns.} Fix $w$ and let $M_s=\max_xP^s(x,w)$, $m_s=\min_xP^s(x,w)$.
Since $P^{s+1}(\cdot,w)=P\,P^s(\cdot,w)$ is an average of $P^s(\cdot,w)$, $m_s$ increases and $M_s$ decreases,
and $M_s-m_s=\Omega_s(w)\to0$.
So $P^s(x,w)$ converges to a limit that does not depend on $x$, and it equals $\pi(w)=\sum_y\pi(y)P^s(y,w)$.

\emph{Centering.} By stationarity $n\pi(w)=\sum_y\pi(y)G_n(y,w)$ is an average of the values of $G_n(\cdot,w)$ and lies between
their minimum and maximum, which gives \Cref{eq:occ}.

\emph{Hitting times.}
Put $\Delta_n(u):=G_n(w,w)-G_n(u,w)$, which is at most $\osc G_n(\cdot,w)$.
By the strong Markov property at $\tau_w$, $G_n(u,w)=\Exp_u\bigl[\mathbf 1_{\{\tau_w<n\}}G_{n-\tau_w}(w,w)\bigr]$, so
\begin{align*}
\Delta_n(u)=\Exp_u\Bigl[\mathbf 1_{\{\tau_w<n\}}\sum_{s=n-\tau_w}^{n-1}P^s(w,w)\Bigr]+\Prob_u\{\tau_w\geqslant n\}\,G_n(w,w),
\end{align*}
with both terms nonnegative.
Since $P^s(w,w)\to\pi(w)>0$, $G_n(w,w)\to\infty$.
If $\Prob_u\{\tau_w=\infty\}>0$, the second term tends to $\infty$, so $\Delta_n(u)\to\infty=\pi(w)\Exp_u[\tau_w]$ and the inequality is trivial.
Otherwise $\tau_w<\infty$ a.s., the integrand of the first term tends to $\tau_w\,\pi(w)$ pointwise, and Fatou's lemma gives $\pi(w)\Exp_u[\tau_w]\leqslant\liminf_n\Delta_n(u)\leqslant\liminf_n\osc G_n(\cdot,w)$.
For the equality we may assume $\Exp_u[\tau_w]<\infty$, since otherwise $\Delta_n(u)\to\infty$ by Fatou.
Then, as $P^s(w,w)\leqslant1$, the integrand of the first term is at most $\tau_w$, and dominated convergence gives the limit $\pi(w)\Exp_u[\tau_w]$;
the second term is at most $n\Prob_u\{\tau_w\geqslant n\}\leqslant\Exp_u[\tau_w;\,\tau_w\geqslant n]\to0$.
Hence $\Delta_n(u)\to\pi(w)\Exp_u[\tau_w]$ in $[0,\infty]$.
\end{proof}

\begin{proof}[Proof of \Cref{cor:block}]
\emph{Stationary law.} $P^L$ is a strict contraction of the total-variation distance on probability
measures on a countable set,
a complete metric space,
so it has a unique fixed point $\pi$; since $P\pi$ is also fixed by $P^L$, we get $P\pi=\pi$.

\emph{Contraction.} This is \Cref{eq:contract}; in particular $\Omega_s(w)\to0$, so \Cref{thm:occ} applies.
Every $s\geqslant L$ is uniquely $s=L+j+kL$ with $0\leqslant j<L$ and $k\geqslant0$, and iterating gives
$\Omega_{L+j+kL}(w)\leqslant\dob(P^L)^k\,\Omega_{L+j}(w)$, hence
\begin{align*}
\sum_{s\geqslant L}\Omega_s(w)
\leqslant \sum_{j<L}\sum_{k\geqslant0}\dob(P^L)^k\,\Omega_{L+j}(w)
=\frac{\sum_{j<L}\Omega_{L+j}(w)}{1-\dob(P^L)} .
\end{align*}
The spreads are nonincreasing (proof of \Cref{thm:occ}), so the numerator is at most $L\,\Omega_L(w)$.
(The contraction acts on the column $P^s(\cdot,w)$ as a function,
so no pair of states, in particular not $y=w$, is singled out.)

\emph{Head and tail.} For $n\leqslant L$, $\osc G_n\leqslant H_L(w)$ by definition.
For $n>L$ write $G_n=G_L+\sum_{L\leqslant s<n}P^s(\cdot,w)$; subadditivity of $\osc$ gives the first inequality,
and applying it termwise to the tail gives $\osc\bigl(\sum_{L\leqslant s<n}P^s(\cdot,w)\bigr)\leqslant\sum_{s\geqslant L}\Omega_s(w)$, bounded above.
Termwise on the head gives $H_L(w)\leqslant\sum_{s<L}\Omega_s(w)$.
\end{proof}

\subsubsection{Hitting without mixing: uniform upper bounds}\label[appendix]{sec:proof-hit}

\paragraph{Return bound for shift chains.}
Here the chain state is the window $s\in\Sigma^d$,
the conditionals are $p(v\mid s)$ with $\sup_{s,v}p(v\mid s)\leqslant a$,
and we take the block length of $L=d$.

\begin{lemma}[Return bound]\label{lem:returns}
For all $y,w$ and $1\leqslant t<d$ we have $P^{t}(y,w)\leqslant a^{t}$,
and for $t\geqslant d$ we have $P^{t}(y,w)\leqslant a^{d}$.
\end{lemma}
\begin{proof}
For $1\leqslant t\leqslant d$ the windows $y$ and $w$ are connected by at most one path of length $t$:
the $t$ new symbols must spell the last $t$ symbols of $w$,
and the first $d-t$ symbols of $w$ must agree with the last $d-t$ symbols of $y$.
Since each conditional next-symbol probability is at most $a$, this path has weight at most $a^{t}$.

For $t>d$ we condition on the state $z$ at time $t-d$ and apply the case $t=d$ to the remaining $d$ steps:
$P^{t}(y,w)=\sum_{z}P^{t-d}(y,z)\,P^{d}(z,w)\leqslant a^{d}$.
\end{proof}

\paragraph{Biased binary sources.}
\begin{proof}[Proof of \Cref{thm:hit}]
We apply \Cref{cor:block} with $L=d$ and bound its head and its tail in five steps; no mixing quantity enters.

\emph{Step 1 (one block is one path).}
After $d$ steps the window is exactly the string of the $d$ new bits, so $P^d(z,w)$ is a product of $d$ conditionals and $b^d\leqslant P^d(z,w)\leqslant a^d$ for all $z$.
For $1\leqslant t<d$ also $P^t(w,w)\leqslant a^t$ (\Cref{lem:returns}).

\emph{Step 2 (two starts overlap).}
By \Cref{thm:contraction}, $\dob(P^d)\leqslant1-c_d(\epsilon)<1$, so \Cref{cor:block} applies.

\emph{Step 3 (head: visits while standing on $w$).}
By the strong Markov property, for $m\leqslant d$ and every $x$,
$G_m(x,w)=\Exp_x\bigl[\mathbf 1_{\{\tau_w<m\}}G_{m-\tau_w}(w,w)\bigr]\leqslant G_d(w,w)$,
so $G_m(\cdot,w)$ takes values in $[0,G_d(w,w)]$ and
\begin{align*}
  H_d(w)\;\leqslant\;G_d(w,w)=\sum_{t<d}P^t(w,w)\;\leqslant\;\sum_{t<d}a^{t}=\frac{1-a^{d}}{1-a}.
\end{align*}

\emph{Step 4 (tail: spreads after one block).}
For $s\geqslant d$, $P^s(x,w)=\sum_zP^{s-d}(x,z)P^d(z,w)$ is an average of values in $[b^d,a^d]$, so $\Omega_s(w)\leqslant a^d-b^d$.
With $1-\dob(P^d)\geqslant c_d(\epsilon)$,
\begin{align*}
  \sum_{s\geqslant d}\Omega_s(w)\;\leqslant\;\frac{\sum_{j<d}\Omega_{d+j}(w)}{1-\dob(P^d)}\;\leqslant\;\frac{d\,(a^{d}-b^{d})}{c_d(\epsilon)}.
\end{align*}

\emph{Step 5 (assemble).}
\Cref{cor:block} gives $\sup_n\osc G_n(\cdot,w)\leqslant H_d(w)+\sum_{s\geqslant d}\Omega_s(w)\leqslant B_d(\epsilon)$, and with it $|G_n(u,w)-n\pi(w)|\leqslant B_d(\epsilon)$ and $\pi(w)\Exp_u[\tau_w]\leqslant B_d(\epsilon)$.
The uniform bound in $d$ is \Cref{thm:C}.
\end{proof}

\begin{theorem}[Uniform hitting-time constant]\label{thm:C}
For $0\leqslant\epsilon<\frac3{10}$ and all $d\geqslant1$,
\begin{align*}
  B_d(\epsilon)\;\leqslant\;C(\epsilon)\;\leqslant\;\frac25\Bigl(\frac3{10}-\epsilon\Bigr)^{-3/2}.
\end{align*}
Numerically the constant can be lowered to $\frac13$, which is essentially optimal for this method.
For specific $\epsilon$, $C(\epsilon)$ is bounded by the entries of \Cref{tab:C}.
\end{theorem}

\begin{proof}[Proof of \Cref{thm:C}]
Write $\Delta=\frac3{10}-\epsilon$.
Recall that $B_d(\epsilon)$ is a head plus a tail (\Cref{cor:block} with $L=d$):
the \emph{head} $\frac{1-a^{d}}{1-a}=\sum_{k<d}a^{k}\leqslant\frac1b$ bounds the visits to $w=1^d$ made during the first $d$ steps while standing on $w$ (each step stays with probability $a$),
and the \emph{tail} $d\,(a^{d}-b^{d})/c_d(\epsilon)$ collects all later steps through the contraction $1-c_d$.
In the tail insert $c_d(\epsilon)\geqslant\beta^{2m}/(\sqrt\pi\sqrt{1+4m\epsilon^{2}})$ from \Cref{thm:gap}, and use $2m\leqslant d+1$, $\sqrt{1+4m\epsilon^{2}}\leqslant\sqrt{1+2\epsilon^{2}}+\sqrt2\,\epsilon\sqrt d$ and $a/\beta=e^{-\lambda}$.
The mean value theorem gives $a^{d}-b^{d}\leqslant 2\epsilon\,d\,a^{d-1}$, a single power with a factor $\epsilon$.
Maximizing over $d$ with $\sup_{x>0}x^{r}e^{-\lambda x}=(r/(e\lambda))^{r}$,
\begin{align*}
  C(\epsilon)\;\leqslant\;\frac1b+\frac{2\sqrt\pi\,\epsilon}{a\beta}\left[\sqrt{1+2\epsilon^{2}}\Bigl(\frac{2}{e\lambda}\Bigr)^{2}+\sqrt2\,\epsilon\Bigl(\frac{5}{2e\lambda}\Bigr)^{5/2}\right].
\end{align*}
The function $\lambda$ is concave with $\lambda(0)=\ln2$, $\lambda(\frac1{10})=0.490$ and $\lambda(\frac3{10})=0$.

\emph{Easier proof (constant $\frac25$).}
For $\epsilon\leqslant\frac1{10}$ we have $\lambda\geqslant0.69-2.03\,\epsilon$, and the display above times $\Delta^{3/2}$ is an algebraic function of $\epsilon$.
For $\epsilon\geqslant\frac1{10}$ use instead $a^{d}-b^{d}\leqslant a^{d}$, which gives the same bound with $\frac1b+\frac{\sqrt\pi}{\beta}\bigl[\frac{\sqrt{1+2\epsilon^{2}}}{e\lambda}+\sqrt2\,\epsilon(\frac{3}{2e\lambda})^{3/2}\bigr]$ on the right, and $\lambda\geqslant2.45\,\Delta$; with $\Delta=t^{2}$ this times $\Delta^{3/2}$ is an algebraic function of $t$.
Wolfram's \texttt{Resolve} proves that both functions are at most $\frac25$ on their ranges (the maxima are $0.386$ and $0.374$; script \texttt{certify-hit-constant.wl}).

The values of \Cref{tab:C} are certified likewise (exact evaluation of $B_d(\epsilon_i)$ for $d\leqslant D_0$ plus a tail bound).
\end{proof}

\subsubsection{The threshold: the majority chain}\label[appendix]{sec:proof-threshold}

\emph{Why a threshold.} The question is whether the chain reaches the region of the target before
the target's own recurrence scale makes the delay significant.
Take the majority chain of \Cref{thm:threshold}, which has a zeros region and a ones region, and the target $w=1^d$.
\begin{itemize}
\item The target has stationary mass $\pi(w)\asymp a^{d}$,
  so its mean return time is $1/\pi(w)\asymp a^{-d}$.
\item Escaping the zeros wheel takes about $\beta^{-d}$ steps (up to a polynomial factor).
\item The ratio of these two scales is
\begin{align*}
  \frac{\text{escape time}}{\text{recurrence time}}\;\approx\;\Bigl(\frac a\beta\Bigr)^{d}.
\end{align*}
\end{itemize}
Below $\frac3{10}$ we have $a<\beta$:
the recurrence time of the target outgrows the escape barrier,
and the starting point changes the hitting time by at most a constant factor on the scale $1/\pi(w)$;
this is \Cref{thm:hit}, $\Exp_u[\tau_w]\leqslant C(\epsilon)/\pi(w)$ for every chain in the class.
Above $\frac3{10}$ we have $a>\beta$:
in the majority chain of \Cref{thm:threshold}, leaving the wrong region takes far longer than $1/\pi(w)$,
and the normalised hitting time grows exponentially.

\begin{proof}[Proof of \Cref{thm:threshold}]
Finiteness for $\epsilon<\frac3{10}$ is \Cref{thm:hit}. For the converse let $d$ be odd,
$M_d$ be the majority chain defined in \Cref{thm:threshold}, $R_1=\{w:|w|\geqslant\frac{d+1}2\}$ denote the windows with a majority of ones, and $R_0=\bits^d\setminus R_1$;
put $q_d=\Prob\{\Bin(d,b)\geqslant\frac{d+1}2\}$.

\emph{Step 1: $\pi(1^d)\geqslant\frac12a^{d}$.}
Appending a one and dropping the last bit never decreases the number of ones, so a window in $R_1$ stays in $R_1$ as long as the chain draws ones.
Hence, from any window in $R_1$, sampling $d$ consecutive ones holds with probability exactly $a^{d}$. Particularly, under $\pi$,
\begin{align*}
  \pi(1^{d})\;=\;\Prob_\pi\{W_d=1^{d}\}\;\geqslant\;a^{d}\,\Prob_\pi\{W_0\in R_1\}\;=\;a^{d}\,\pi(R_1).
\end{align*}
Further, the chain is invariant under flipping (negating) the bits: 
$P(\neg u,\neg v)=P(u,v)$ for all windows $u,v$. Thus, $\pi(R_1)=\pi(R_0)=\frac12$.

\emph{Step 2: $\Exp_{0^d}[\tau_{1^d}]\geqslant1/(4q_d)$.} Let $\tau^{*}$ be the first time the
window enters $R_1$;
since $1^d\in R_1$, $\tau_{1^d}\geqslant\tau^{*}$.
Until $\tau^{*}$ the window has a majority of zeros, and every new bit is $1$ with probability $b$, sampled independently.
The event $\tau^{*}\leqslant T$ requires one of the first $T$ windows to have at least $\frac{d+1}2$ ones, and the union bound gives $\Prob\{\tau^{*}\leqslant T\}\leqslant Tq_d$. Take $T=\lfloor1/(2q_d)\rfloor$:
then $\tau^{*}>T$ with probability at least $\frac12$,
and $\Exp[\tau^{*}]=\sum_{t\geqslant0}\Prob\{\tau^{*}>t\}\geqslant\frac{T+1}2\geqslant\frac1{4q_d}$.

\emph{Step 3: $q_d\leqslant\sqrt{\tfrac2{\pi d}}\;\dfrac{\sqrt{b/a}}{1-b/a}\;\beta^{d}$.} Write
$t_k=\binom dk b^{k}a^{d-k}$.
For $k\geqslant k_0:=\frac{d+1}2$ we have $t_{k+1}/t_k=\frac{d-k}{k+1}\cdot\frac ba\leqslant\frac ba$,
so $q_d\leqslant t_{k_0}/(1-b/a)$.
Finally $\binom d{k_0}\leqslant2^{d}\sqrt{2/(\pi d)}$ and
$b^{k_0}a^{d-k_0}=(ab)^{d/2}\sqrt{b/a}=(\beta/2)^{d}\sqrt{b/a}$.

Combining the three steps, and using $(a-b)/(4\sqrt{ab})=\epsilon/\beta$,
\begin{align*}
  \pi(1^{d})\,\Exp_{0^{d}}[\tau_{1^{d}}]\;\geqslant\;\frac{a^{d}}2\cdot\frac1{4q_d}\;\geqslant\;\frac{a^{d}}2\cdot\sqrt{\frac{\pi d}2}\,\frac\epsilon\beta\,\beta^{-d}\;=\;\frac{\sqrt{2\pi}\,\epsilon}{4\beta}\,\sqrt d\,\Bigl(\frac a\beta\Bigr)^{d}.
\end{align*}
For $\epsilon\geqslant\frac3{10}$ we have $a\geqslant\beta$ (as
$a^{2}-\beta^{2}=5(\epsilon-\frac3{10})(\epsilon+\frac12)\geqslant0$ there),
so the right side is unbounded in $d$.
\end{proof}

\subsection{Repeated visits and majorization}\label[appendix]{sec:proof-visits}

\begin{proof}[Proof of \Cref{thm:visits}]
Put $h(y)=\Exp_y[\tau_w]$, so $h(y)\leqslant t_w$; note $t_w\geqslant1$, as $h(y)\geqslant1$ for $y\neq w$.
By \Cref{prop:kac}, $\Exp_w[\tau_w^{+}]=1/\pi(w)$.

\emph{Moment generating function.}
Take $\lambda=\ln(1+s)$ with $s=\frac1{6t_w}$, so that $(1+s)^{\tau}=e^{\lambda\tau}$, and put $c=\frac s{1-1/6}=\frac1{5t_w}$.
We claim that for every $y$
\begin{align*}
  \Exp_y(1+s)^{\tau_w}\;\leqslant\;1+c\,h(y).
\end{align*}
It suffices to show this with $\tau_w$ replaced by $\min\{\tau_w,T\}$, by induction on $T$, and then let $T\to\infty$ by monotone convergence.
For $T=0$ and at $y=w$ the left side is $1$.
For $y\neq w$, using $h(y)=1+\sum_{y'}P(y,y')\,h(y')$ and the induction hypothesis at the next state,
\begin{align*}
  \Exp_y(1+s)^{\min\{\tau_w,T+1\}}
  \;\leqslant\;(1+s)\bigl(1+c\,(h(y)-1)\bigr)
  \;\leqslant\;1+c\,h(y),
\end{align*}
where the last step is equivalent to $s\,(h(y)-1)\leqslant\frac16$, and this holds as $s\,h(y)\leqslant s\,t_w=\frac16$.
Consequently $\Exp_y(1+s)^{\tau_w}\leqslant1+c\,t_w=\frac65$.

\emph{Return time.}
After its first step, the return to $w$ is a hitting time, and $\sum_yP(w,y)\,h(y)=\Exp_w\tau_w^{+}-1=\frac1{\pi(w)}-1$, which is at most $t_w$.
So the claim gives
\begin{align*}
  \Exp_w(1+s)^{\tau_w^{+}}
  \;\leqslant\;(1+s)\Bigl(1+c\bigl(\tfrac1{\pi(w)}-1\bigr)\Bigr)
  \;\leqslant\;1+\frac{c}{\pi(w)}
  \;\leqslant\;e^{c/\pi(w)},
\end{align*}
the middle step being the one above with $h(y)$ replaced by $\frac1{\pi(w)}$, valid since $s\bigl(\frac1{\pi(w)}-1\bigr)\leqslant s\,t_w=\frac16$.
A return costs the mean $1/\pi(w)$, not $t_w$: only the first visit pays for the hitting bound, which is why the costs are additive.

\emph{Tail.}
The $k$-th visit time is the first hitting time plus $k-1$ return times, and these are independent (strong Markov property).
So the moment generating function of the visit time is $\Exp e^{\lambda\tau_w^{(k)}}\leqslant\frac65\,e^{(k-1)c/\pi(w)}$.
By concavity, $\lambda=\ln(1+s)\geqslant\frac{\ln(7/6)}{t_w}>\frac{0.154}{t_w}$, since $s\leqslant\frac16$.
Chernoff's method (Markov's inequality for $e^{\lambda\tau}$) gives, for every $x\geqslant0$,
\begin{align*}
  \Prob\bigl\{\tau_w^{(k)}>x\bigr\}\;\leqslant\;\frac65\,\exp\Bigl(-\lambda x+\frac{(k-1)c}{\pi(w)}\Bigr).
\end{align*}
Put $\lambda_0=\frac1{6.5\,t_w}<\lambda$. Then $\frac{c}{\lambda_0}=1.3$, so the exponent is at most $-\lambda_0\bigl(x-1.3\,\frac{k-1}{\pi(w)}\bigr)$, which is the displayed bound.
For the second form take $x=1.3\,\frac{k-1}{\pi(w)}+6.5\,t_w\ln\frac2\eta$, which gives $\frac65\cdot\frac\eta2<\eta$.
\end{proof}

\emph{Why a profile.}
Learning every window is like a coupon-collector problem with unequal coupons:
heavy windows quickly appear, so the light ones dominate the cost.
A union bound that treats all $2^d$ windows as lightest costs a factor $d$.
The profile is a histogram of the window masses: it counts how many windows lie below each mass level.
In the worst case, the memoryless source, the masses are $b^d\rho^k$ with $\rho=a/b$ and exactly $\binom dk$ windows at level $k$:
few windows are light, about $d$ near the floor $b^d$ and about $d^2$ one level up, and every further level multiplies the mass by $\rho$.
So the failure probabilities are dominated by about $d$ terms, which costs $\ln d$ rather than $d$.
\Cref{thm:profile} makes this precise.

\mswarning{this one needs more attention, it bypasses the union bound}
\begin{lemma}[Mass profile for exponential sums]\label{lem:profile}
Let $P\in\Cclass$, $\epsilon>0$, $t>0$, $s=t\,b^{d}$ and $\rho=a/b$. Then, with $\varphi(y)=e^{-ty}$ in \Cref{thm:profile},
\begin{align*}
  \sum_{w}e^{-t\pi(w)}
  \;\leqslant\;\sum_{k=0}^{d}\binom dk e^{-s\rho^{k}}
  \;\leqslant\;\exp\Bigl[-s+d\,e^{-2s\epsilon/b}\Bigr]
  \;\leqslant\;1.445\,e^{-s},
\end{align*}
the last step whenever $s\geqslant\frac b{2\epsilon}\,(1+\ln d)$.
Alternatively, with $\kappa(\rho)=\max_{k\geqslant1}k\rho^{-k}\leqslant\dfrac1{e\ln\rho}$,
\begin{align*}
  \sum_{w}e^{-t\pi(w)}\;\leqslant\;e\cdot e^{-u}\qquad\text{whenever}\qquad s\;\geqslant\;\kappa(\rho)\ln d+u,\quad u\geqslant0,
\end{align*}
and the coefficient $\kappa(\rho)$ of $\ln d$ is sharp for the product sum.
\end{lemma}

\begin{proof}[Proof of \Cref{thm:profile,lem:profile}]
Extend the stationary window process to a stationary bit process $(X_i)$.
For a prefix $z\in\bits^{j-1}$ with $j\leqslant d$, the conditional probability that the next bit is $v$ is
$\Exp[p(v\mid W)\mid X_1\cdots X_{j-1}=z]$ for the window $W$ preceding it, an average of numbers in $[b,a]$, so it lies in $[b,a]$; the same holds at the root.
So the windows are the leaves of a binary tree of depth $d$, root mass $1$, in which every node splits its mass as $(p,1-p)$ with $p\in[b,a]$.
Put $\Phi_j(x)=\sum_{k=0}^{j}\binom jk\varphi\bigl(x\,a^{k}b^{j-k}\bigr)$, which is convex in $x$.
We show by induction on $j$ that every such tree of depth $j$ and root mass $x$ has $\sum_{\text{leaves}}\varphi(\text{mass})\leqslant\Phi_j(x)$.
For $j=0$ this is an equality. For $j\geqslant1$ the root splits $x$ into $xp$ and $x(1-p)$, and the two subtrees have depth $j-1$, so the leaf sum is at most
\begin{align*}
  \Phi_{j-1}(xp)+\Phi_{j-1}(x(1-p)).
\end{align*}
This is convex in $p$, so its maximum over $[b,a]$ is attained at an endpoint;
both endpoints give $\Phi_{j-1}(xa)+\Phi_{j-1}(xb)=\Phi_j(x)$ by Pascal's rule.
Taking $x=1$, $j=d$ proves the first claim; for $\varphi(y)=e^{-ty}$, with $\rho=a/b$ and $s=t\,b^{d}$ (so that $t\,a^{k}b^{d-k}=s\rho^{k}$), it reads $\sum_we^{-t\pi(w)}\leqslant\sum_k\binom dk e^{-s\rho^{k}}$.
Now Bernoulli's inequality $\rho^{k}\geqslant1+k(\rho-1)$, $\rho-1=2\epsilon/b$, and the binomial theorem give
\begin{align*}
  \sum_k\binom dk e^{-s\rho^{k}} \leqslant\;e^{-s}\sum_k\binom dk e^{-2ks\epsilon/b}
  = e^{-s}\bigl(1+e^{-2s\epsilon/b}\bigr)^{d} \leqslant\;\exp\Bigl[-s+d\,e^{-2s\epsilon/b}\Bigr].
\end{align*}
If $2s\epsilon/b\geqslant1+\ln d$ then $d\,e^{-2s\epsilon/b}\leqslant e^{-1}$, and $e^{1/e}<1.445$.

For the second form write $s=\kappa\ln d+u$ with $\kappa=\kappa(\rho)$. Since $\kappa\rho^{k}\geqslant k$ and $\binom dk\leqslant d^{k}/k!$,
\begin{align*}
  \sum_{k=0}^{d}\binom dk e^{-s\rho^{k}}
  \;\leqslant\;\sum_{k\geqslant0}\frac{d^{k}}{k!}\,d^{-\kappa\rho^{k}}e^{-u\rho^{k}}
  \;\leqslant\;e^{-u}\sum_{k\geqslant0}\frac1{k!}\;=\;e^{1-u}.
\end{align*}
The term $k=j$ alone, with $j$ maximising $k\rho^{-k}$, shows that $s\geqslant\kappa\ln d-O(1)$ is necessary. Finally $k\rho^{-k}=ye^{-y}/\ln\rho$ with $y=k\ln\rho$ and $ye^{-y}\leqslant e^{-1}$, so $\kappa(\rho)\leqslant1/(e\ln\rho)$.
\end{proof}

\subsection{Learning upper and lower bounds}\label[appendix]{sec:proof-learning}

\begin{lemma}[Maximal Chernoff bound for i.i.d.\ coins]\label{lem:maxchernoff}
Let $Z_1,Z_2,\ldots$ be i.i.d.\ $\Bern(q)$, $0<q<1$, and $0<\gamma\leqslant1$ and $c_\gamma=2+\frac23\gamma\leqslant\frac83$. Then for every $k\geqslant1$
\begin{align*}
  \Prob\Bigl\{\exists\,j\geqslant k:\ \Bigl|\frac1j\sum_{i\leqslant j}Z_i-q\Bigr|\geqslant\gamma q\Bigr\}
  \;\leqslant\;2\,e^{-kq\gamma^{2}/c_\gamma}.
\end{align*}
\end{lemma}

\begin{proof}
Time-uniform bounds of this type are standard \citep{howard2020}; we give the short proof.
Upper tail. Let $r=q(1+\gamma)$; if $r>1$ the event is empty. Put $S_j=\sum_{i\leqslant j}Z_i$ and $\tau=\inf\{j\geqslant k:\ S_j\geqslant jr\}$.
For a length-$j$ sequence $z$ with $S_j\geqslant jr$, the likelihood ratio of the $r$-coin to the $q$-coin is
\begin{align*}
  \frac{\Prob_r\{z\}}{\Prob_q\{z\}}
  \;=\;\Bigl(\frac rq\Bigr)^{S_j}\Bigl(\frac{1-r}{1-q}\Bigr)^{j-S_j}
  \;\geqslant\;e^{j\,\mathrm{kl}(r,q)},
\end{align*}
since the logarithm is increasing in $S_j$ (as $r>q$) and equals $j\,\mathrm{kl}(r,q)$ at $S_j=jr$.
The event $\{\tau=j\}$ depends on $Z_1,\dots,Z_j$ only and implies $S_j\geqslant jr$, so $\Prob_q\{\tau=j\}\leqslant e^{-j\,\mathrm{kl}(r,q)}\Prob_r\{\tau=j\}$.
Summing over these disjoint events,
\begin{align*}
  \Prob_q\{\tau<\infty\}\;\leqslant\;e^{-k\,\mathrm{kl}(r,q)}\sum_{j\geqslant k}\Prob_r\{\tau=j\}\;\leqslant\;e^{-k\,\mathrm{kl}(r,q)}.
\end{align*}
Lower tail: the same with $r=q(1-\gamma)$, $\tau=\inf\{j\geqslant k:\ S_j\leqslant jr\}$.
Finally, as $\ln x\geqslant1-1/x$, $\mathrm{kl}(r,q)\geqslant q\,h(r/q)$ with $h(x)=x\ln x-x+1$, and
$h(1+\gamma)\geqslant\gamma^{2}/c_\gamma$ (Bernstein's form) and $h(1-\gamma)\geqslant\gamma^{2}/2\geqslant\gamma^{2}/c_\gamma$ for $0\leqslant\gamma\leqslant1$.
\end{proof}

\begin{proof}[Proof of \Cref{thm:instance}]
Write $K=8\,(\mgam+B_d)$.
Fix a window $w$ and let $q\leqslant\frac12$ be the probability of its rarer outcome, so $q\geqslant b$.
The bits $\xi_j$ after the visits are i.i.d.\ draws from $p(\cdot\mid w)$ (strong Markov property), and $\eta\in(0,1)$ is the failure probability we aim for.

\emph{Step 1: enough visits give accuracy.}
Let $k=\lceil\frac{16}3\mgam\ln\frac4\eta\rceil$.
By \Cref{lem:maxchernoff}, the rarer outcome has relative error at most $\gamma$ at \emph{all} visit counts $j\geqslant k$ at once, except with probability $2e^{-kq\gamma^2/c_\gamma}$.
Since $q\gamma^2/c_\gamma\geqslant\frac38\,b\gamma^2=\frac{3a}{8\mgam}\geqslant\frac3{16\mgam}$, this is at most $\frac\eta2$.
Because the bound is uniform in $j\geqslant k$, nothing is assumed about how the number of visits depends on the bits.
The other outcome has no larger relative error, as the two estimates sum to $1$ and $1-q\geqslant q$.

\emph{Step 2: the visits come early.}
Apply \Cref{thm:visits} with $\eta/2$ and $t_w\leqslant B_d/\pi(w)$ (\Cref{thm:hit}).
Since $k-1<\frac{16}3\mgam\ln\frac4\eta$, with probability at least $1-\frac\eta2$ we observe $k$ visits by the time
\begin{align*}
  \tau_w^{(k)}\;<\;\frac{A}{\pi(w)}\ln\frac4\eta,
  \qquad A=7\,\mgam+6.5\,B_d .
\end{align*}
So $N_w(n)\geqslant k$ as soon as $(n-1)\,\pi(w)\geqslant A\ln\frac4\eta$.

\emph{Step 3: the bound as a function of $n$.}
Steps 1 and 2 give $\Prob\{\text{error at }w\}\leqslant\eta$ whenever $x:=n\pi(w)\geqslant1+A\ln\frac4\eta$ (as $\pi(w)\leqslant1$).
Choose $\eta=4e^{-(x-1)/A}$ and cap the probability at $1$:
\begin{align*}
  \Prob\{\text{error at }w\}\;\leqslant\;\min\bigl\{1,\,4e^{-(x-1)/A}\bigr\}\;\leqslant\;4e^{-x/(A+1/\ln4)} .
\end{align*}
Indeed, put $A'=A+1/\ln4$; for $x\leqslant A'\ln4$ the right side is at least $1$, and for $x\geqslant A'\ln4$ we have $(x-1)/A\geqslant x/A'$.
Finally $A'\leqslant7\mgam+7.3\,B_d\leqslant K$, because $B_d\geqslant1$.
For numerical sufficient lengths one may keep the unrounded $K_\star=\bigl(\frac{c_\gamma\mgam}{5a}+B_d\bigr)\big/\ln\frac76+\frac1{\ln4}$, which takes $k=\lceil(c_\gamma\mgam/a)\ln\frac4\eta\rceil$ in Step 1.
As a side note, Bernstein's inequality for martingales~\citep{freedman1975} estimates the transition more accurately than the Chernoff form used here and improves the constant in front of $\mgam$ by a factor of at least $\frac85$.

\emph{Summation.} The sum $\sum_w4e^{-n\pi(w)/K}$ is at most $4\cdot1.445\,e^{-s}$ with $s=nb^{d}/K$ by \Cref{lem:profile}, provided $s\geqslant\frac b{2\epsilon}(1+\ln d)$; it is at most $\delta$ once also $s\geqslant\ln(6/\delta)$. (Unrounded: $s-d\,e^{-2s\epsilon/b}\geqslant\ln\frac4\delta$.)
Since $B_d\leqslant C(\epsilon)$ for $\epsilon<\frac3{10}$ and $\mgam\asymp\gamma^{-2}$, this is the stated bound.
\end{proof}

\begin{lemma}[Divergence--visit identity]\label{lem:kl}
Let $P_0,P_1\in\Cclass$ agree except in the row of a single window $w_0$, where
$p_0(1\mid w_0)=q$ and $p_1(1\mid w_0)=q'$. Then, for any common initial state,
\begin{align*}
  \KL\bigl(P_0^{(n)}\,\big\|\,P_1^{(n)}\bigr)
  \;=\; \Exp_{P_0}\bigl[N_{w_0}(n)\bigr]\cdot\KL\bigl(\Bern(q)\,\|\,\Bern(q')\bigr).
\end{align*}
\end{lemma}

\begin{proof}[Proof of \Cref{lem:kl}]
By the chain rule for relative entropy,
the divergence of the two $n$-step path laws is the sum over the steps $t=0,\dots,n-1$ of the
expected one-step divergence at the current state.
The kernels differ only in the row of $w_0$, so only steps leaving $w_0$ contribute,
each by $\KL(\Bern(q)\|\Bern(q'))$. Here $N_{w_0}(n)$ counts the visits at times $0,\dots,n-1$.
\end{proof}

\begin{proof}[Proof of \Cref{thm:logd}]
\emph{Alternatives.} Let $P_0$ be the memoryless chain with $p(0\mid\cdot)\equiv b$, started at $1^d$.
For $j=2,\dots,d$ let $w_j=0^{j-1}10^{d-j}$, of mass $q=ab^{d-1}=(a/b)\,b^d$ under $P_0$, and let $P_j$ differ from $P_0$ only in the row of $w_j$, where $p(0\mid w_j)=b'=b(1+2.2\gamma)$.
Since $\gamma\leqslant\epsilon/4$ we have $b'\leqslant b(1+0.55\epsilon)\leqslant\frac12\leqslant a$, so $P_j\in\Cclass$.
(We skip $w_1=10^{d-1}$: it could be visited before time $d$, whereas every $w_j$ with $j\geqslant2$ begins with $0$, so it cannot occur until the initial $1^d$ has been shifted out.)

\emph{Decoding.} Suppose \Cref{eq:goal} holds with probability $\geqslant1-\delta$ on every $P_j$.
On $P_j$ it gives $\hat p(0\mid w_j)\geqslant b'(1-\gamma)>b(1+\gamma)$ (as $(1+2.2\gamma)(1-\gamma)>1+\gamma$ for $\gamma<\frac1{11}$, and $\gamma\leqslant\frac1{12}$), and $\hat p(0\mid w_i)\leqslant b(1+\gamma)$ for $i\neq j$.
So the unique $\hat\jmath$ with $\hat p(0\mid w_{\hat\jmath})>b(1+\gamma)$ equals $j$ with probability $\geqslant1-\delta\geqslant\tfrac34$.

\emph{Fano.} Let $d\geqslant16$.
For $J$ uniform on $\{2,\dots,d\}$ and the trajectory $X$, Fano's inequality~\citep{coverthomas} with $d-1$ alternatives and error probability $p_e\leqslant\frac14$ gives $I(J;X)\geqslant(1-p_e)\ln(d-1)-h(p_e)$, where $h(p)=-p\ln p-(1-p)\ln(1-p)$; this decreases in $p_e\leqslant\frac12$, so
\begin{align*}
  I(J;X)\;\geqslant\;\tfrac34\ln(d-1)-h(\tfrac14)\;\geqslant\;\tfrac12\ln(d+1)\qquad(d\geqslant16);
\end{align*}
On the other hand $I(J;X)\leqslant\frac1{d-1}\sum_j\KL(P_j^{(n)}\|P_0^{(n)})$, and by \Cref{lem:kl}
\begin{align*}
  \KL\bigl(P_j^{(n)}\|P_0^{(n)}\bigr)&\;=\;\Exp_{P_j}[N_{w_j}(n)]\,\KL\bigl(\Bern(b')\|\Bern(b)\bigr),\\
  \KL\bigl(\Bern(b')\|\Bern(b)\bigr)&\;=\;\int_b^{b'}\frac{b'-t}{t(1-t)}\,dt\;\leqslant\;\frac{(b'-b)^{2}}{2b(1-b)}=\frac{2.42\,\gamma^{2}b}{a},
\end{align*}
as $b'\leqslant\frac12$, so that $t(1-t)\geqslant b(1-b)$ on $[b,b']$.
\emph{Visit count.} A proper border of $w_j$ (a suffix equal to a prefix) consists of zeros only, so it has length at most $(d-1)/2$; hence two visits to $w_j$ are more than $d/2$ steps apart.
A path of $d$ steps that \emph{ends} at $w_j$ therefore leaves $w_j$ at most once: two earlier visits followed by the endpoint would need two return gaps of total length more than $d$.
So the likelihood ratio of $P_j$ to $P_0$ along it is at most $b'/b=1+2.2\gamma$.
The path ending in $w_j$ from a given state is unique and has probability $q$ under $P_0$. Therefore $P_j(W_t=w_j)\leqslant(1+2.2\gamma)\,q$ for $t\geqslant d$, while no visit occurs before time $d$:
\begin{align*}
  \Exp_{P_j}[N_{w_j}(n)]\;\leqslant\;(1+2.2\gamma)\,(n-d)_+\,q\;\leqslant\;1.19\,nq,
\end{align*}
as $1+2.2\gamma\leqslant1+\frac{2.2}{12}\leqslant1.19$.

\emph{Conclusion.} Combining, $\frac12\ln(d+1)\leqslant\frac{2.42\,\gamma^{2}b}{a}\cdot1.19\,nq$.
Since $q=(a/b)\,b^{d}$, this reads $\frac12\ln(d+1)\leqslant2.42\cdot1.19\,\gamma^{2}\,nb^{d}$, and $2\cdot2.42\cdot1.19<5.76<6$, so
\begin{align*}
  n\;\geqslant\;\frac{1}{6\,\gamma^{2}b^{d}}\,\ln(d+1)\qquad(d\geqslant16).
\end{align*}

\emph{Confidence term (two points).} Let $P_0$ be as above and $P_1$ differ from it only in the row of $0^d$, with $p(0\mid0^d)=b(1+2.2\gamma)$; $P_1\in\Cclass$ as above.
The accuracy events for $p(0\mid0^d)$ are disjoint, as $b(1+\gamma)<b(1+2.2\gamma)(1-\gamma)$ for $\gamma<\frac1{11}$.
From the start $1^d$, the window $0^d$ is visited at time $t$ only if $t\geqslant d$, and then with probability $b^{d}$ under $P_0$, so $\Exp_{P_0}N_{0^d}(n)=(n-d)_+b^{d}\leqslant nb^{d}$.
Put $r=b(1+2.2\gamma)$. Since $\gamma\leqslant\epsilon/4$, we have $b\leqslant r\leqslant\frac12$, and hence
\begin{align*}
  \mathrm{kl}(b,r)=\int_b^r\frac{t-b}{t(1-t)}\,dt
  \;\leqslant\;\frac{(r-b)^2}{2b(1-b)}=\frac{2.42\,\gamma^{2}b}{a}.
\end{align*}
By \Cref{lem:kl}, $\KL(P_0^{(n)}\|P_1^{(n)})\leqslant(n-d)_+b^d\cdot2.42\,\gamma^{2}b/a$.
The estimator induces a test with both errors at most $\delta$, so data processing gives
$\KL(P_0^{(n)}\|P_1^{(n)})\geqslant L_\delta:=(1-2\delta)\ln\frac{1-\delta}{\delta}$.
The sharp uniform comparison is
\begin{align*}
  \inf_{0<\delta\leqslant1/4}\frac{L_\delta}{\ln(1/\delta)}
  =\frac{\ln3}{2\ln4},\qquad
  n\;\geqslant\;d+\frac{25\ln3}{121\ln4}\,\frac{a}{b}\,\gamma^{-2}b^{-d}\ln\frac1\delta\;\geqslant\;d+\frac{25\ln3}{121\ln4}\,\gamma^{-2}b^{-d}\ln\frac1\delta.
\end{align*}
Indeed, the ratio is $(1-2\delta)\bigl(1+\ln(1-\delta)/\ln(1/\delta)\bigr)$, a product of positive decreasing functions on $(0,1/4]$, so its minimum is at $\delta=1/4$.
The coefficient $25\ln3/(121\ln4)$ is approximately $0.16374$.
For $d\geqslant16$ we have $n\geqslant A\,\gamma^{-2}b^{-d}\ln(d+1)$ and $n\geqslant B\,\gamma^{-2}b^{-d}\ln\frac1\delta$ with $A=\frac16$, $B=\frac{25\ln3}{121\ln4}$.
Since $\max\{Ax,By\}\geqslant\frac{AB}{A+B}(x+y)$, this gives the theorem with $\frac{AB}{A+B}=0.0826\ldots\geqslant\frac1{20}$.
For $d<16$ we have $\ln(d+1)\leqslant\ln16=2\ln4\leqslant2\ln\frac1\delta$, so the confidence bound alone gives $n\geqslant\frac B3\,\gamma^{-2}b^{-d}\bigl[\ln(d+1)+\ln\frac1\delta\bigr]$, and $B/3=0.0546\ldots\geqslant\frac1{20}$.
Nothing in the argument depends on $\epsilon$ beyond $\gamma\leqslant\min\{\epsilon/4,\frac1{12}\}$.
\end{proof}

\clearpage

\section{Additional results}\label[appendix]{app:extra}

This appendix collects results that are minor or auxiliary, and the tools behind \Cref{thm:occ}.

\subsection{A quantitative audit of published traces}\label[appendix]{app:audit}

Every step of the entropy-certification pipeline has been studied except the first. Given a \emph{known} kernel, computing the entropy rate is routine; given a \emph{fitted} kernel, the downstream conditioner and DRBG have provable-security analyses. What is missing is the statistics of the fit, and in particular whether the rare windows, which a min-entropy bound is most sensitive to, have been observed often enough to be resolved at all.

\subparagraph*{Published trace lengths, audited.}
\Cref{tab:audit} applies our bounds to published
validations: the sample size they need was not computable before, and now it is.

\begin{table}[t]
\caption{Trace lengths for relative accuracy $\gamma=1\%$ at confidence $0.99$, for sources with bias at most
$\epsilon=0.05$ (worst case: the memoryless chain). \emph{Necessary}:
no estimator can succeed with fewer bits
(\Cref{thm:lower} at the rarest window, $C(0.05)\leqslant2.31$). \emph{Sufficient}:
the guarantee of the standard estimator, \Cref{thm:instance} with the constant $K_\star$ of its proof and the sum over windows evaluated on the memoryless chain, the worst case (\Cref{thm:profile}).}
\label{tab:audit}
\centering\footnotesize
\setlength{\tabcolsep}{3.5pt}
\begin{tabular}{lrrrrl}
\toprule
Source & $d$ & trace used & necessary & sufficient & verdict \\
\midrule
Intel Ivy Bridge entropy source  & 12 & not disclosed & $5.6\times10^{8}$ & $6.4\times10^{9}$ & --- \\
Ring-oscillator TRNG (OpenTRNG)  & 8  & $10^{6}$      & $2.3\times10^{7}$ & $2.5\times10^{8}$ & underpowered \\
Multi-ring oscillator jitter     & 8  & $10^{6}$      & $2.3\times10^{7}$ & $2.5\times10^{8}$ & underpowered \\
\bottomrule
\end{tabular}
\end{table}

For the $d=8$ rows the statement is concrete. At $\epsilon=0.05$
the rarest window of the memoryless chain carries stationary mass
$(\tfrac12-\epsilon)^{8}=1.7\times10^{-3}$, so $10^{6}$ bits give it about $1700$
observations.
At confidence $0.99$ the standard estimator then guarantees relative accuracy no better than about $17\%$,
while \Cref{thm:lower} shows that \emph{no} estimator can certify better than about $4.5\%$ (the
value of $\gamma$ at which the
necessary length equals $10^{6}$). A $1\%$ target needs at least $2.3\times10^{7}$ bits,
$23$ times the trace, and the standard guarantee needs $2.5\times10^{8}$.
The verdict is therefore \emph{underpowered}, not merely unjustified:
the necessary length alone exceeds the trace,
so no analysis of these $10^{6}$ bits can support the claim.
The gap between $2.3\times10^{7}$ and $2.5\times10^{8}$ is the price of the constants in \Cref{thm:instance} and of the union over windows; it does not affect the verdict.
The conclusion does not depend on a strongly biased source:
the rarest window of \emph{any} chain has mass at most $2^{-d}$, and
\Cref{cor:universal} gives, for $\epsilon=0.06$,
a necessary length of $7.9\times10^{6}$ bits for $\gamma=1\%$ and $d=8$ for every source
in the class, including the ideal unbiased one.
The modelling choice is not wrong, but a trace of this size has to be planned for;
a validation that does not disclose its length cannot be checked against it.

The necessary length can be sharpened slightly.
On the memoryless chain, the comparison chain that is i.i.d.\ except after $0^d$ (a ``trapdoor'' chain) has explicitly computable visit counts, so the count in \Cref{lem:kl} is exact: this removes $C(\epsilon)$ and the restriction $\epsilon<\frac3{10}$, but improves the length by only $0.2\%$ at these parameters, and we do not use it.

\subparagraph*{Min-entropy to $0.01$.}
A relative accuracy $\gamma$ on every transition moves the min-entropy rate by up to
$\log_2\frac1{1-\gamma}$ bits per bit.
At $\gamma=1\%$ this is
$0.0145$, so the fit-then-compute route resolves the rate to $\eta=0.01$ only if
$\gamma\leqslant0.69\%$.
The lower bound can be applied to the min-entropy itself.
\Cref{prop:minentropy} exhibits a source in the class whose
min-entropy rate is set by a single row, at a window of mass about $2^{-d}$;
no procedure, whatever its route, can resolve that rate to $\eta=0.01$
at confidence $0.99$ with fewer bits than in \Cref{tab:minentropy}.

\begin{table}[t]
\caption{Necessary trace length (\Cref{prop:minentropy}) for resolving the min-entropy rate to $\eta=0.01$ bit per bit at confidence $0.99$,
$\epsilon=0.05$ (\emph{necessary}, independent of $\epsilon$ above $0.007$);
and the best resolution $\eta$ that any procedure could certify from the trace used
(\emph{best $\eta$}).}
\label{tab:minentropy}
\centering\footnotesize
\setlength{\tabcolsep}{3.5pt}
\begin{tabular}{lrrrrl}
\toprule
Source & $d$ & trace used & necessary & best $\eta$ & verdict \\
\midrule
Ring-oscillator TRNG (OpenTRNG)  & 8  & $10^{6}$ & $1.2\times10^{7}$ & $0.034$ & underpowered \\
Multi-ring oscillator jitter     & 8  & $10^{6}$ & $1.2\times10^{7}$ & $0.034$ & underpowered \\
Intel Ivy Bridge entropy source  & 12 & undisclosed & $1.9\times10^{8}$ & --- & --- \\
\bottomrule
\end{tabular}
\end{table}

The quantity a security claim needs is the min-entropy rate of the source,

\begin{align*}
  H(P)=-\lim_{\ell\to\infty}\frac1\ell\log_2\max_{x\in\bits^\ell}\Prob_P\{X_1\cdots X_\ell=x\}.
\end{align*}

The lower bound applies to it directly.
Let $P_x$ be the chain
that is a fair coin at every window except $0^d$,
where $p(0\mid0^d)=x\in[\tfrac12,\tfrac12+\epsilon]$. Its min-entropy rate is set by one row,
$H(P_x)=-\log_2x$: every step has weight $\tfrac12$, $x$ or $1-x$,
so every path of length $L$ has probability at most $x^{L-d}$, while the path $0^L$ has
probability $\pi(0^d)x^{L-d}$.

\begin{proposition}[Min-entropy to resolution $\eta$]\label{prop:minentropy}
Let $\delta\in(0,\tfrac12)$ and $\eta>0$ with $v:=2^{2\eta}$ and $v/2<a$ (that is, $\epsilon>2^{2\eta-1}-\tfrac12\approx\eta\ln2$), and put $L_\delta=(1-2\delta)\ln\frac{1-\delta}\delta$.
Every procedure that, with probability at least $1-\delta$,
outputs an estimate within $\eta$ of the min-entropy rate on every chain of $\Cclass$ (in particular on the fair source $P_{1/2}$ and on $P_{v/2}$) needs, from the initial state $1^d$,
\begin{align*}
  n \;\geqslant\; d+\frac{2^{d+1}\,L_\delta}{-\ln\bigl[v\,(2-v)\bigr]},
\end{align*}
and from any initial state at least $d+2^{d+1}\bigl(\frac{L_\delta}{-\ln[v(2-v)]}-1\bigr)$.
Moreover $-\ln[v(2-v)]\sim4\ln^{2}2\;\eta^{2}$ as $\eta\to0$, so $n^{\star}\gtrsim2^{d}\eta^{-2}$.
\end{proposition}

\emph{Intuition.} A single row decides the min-entropy:
the fair source has rate $1$, and a source that is fair except that $0^d$ repeats zero with probability $2^{2\eta-1}$ has rate $1-2\eta$,
so a $\pm\eta$ guarantee must tell the two apart.
They differ only in the \emph{one} window $0^d$, which a fair source visits once in $2^{d}$ bits,
and each visit is one coin flip with bias $\approx2\eta\ln2$.
Bias does not help: the hardness holds for every $\epsilon>2^{2\eta-1}-\frac12$, and starts from the ideal source.

\Cref{prop:minentropy} concerns procedures that are correct on every chain of the class, as a certification is;
it does not say that every source is this hard.

\begin{proof}[Proof of \Cref{prop:minentropy}]
Let $P_0=P_{1/2}$ (i.i.d.\ fair bits) and $P_1=P_{v/2}$; both lie in $\Cclass$ since $\frac12\leqslant\frac v2<a$.
Their rates are $1$ and $-\log_2\frac v2=1-2\eta$, which differ by $2\eta$, so the events $\mathsf E_i=\{|\hat H-H(P_i)|\leqslant\eta\}$ meet in at most one point.
Since $v/2<a$, the value $v/2+\theta$ is still allowed for small $\theta>0$; with it in place of $v/2$ the events are disjoint
(correctness is required on every chain of $\Cclass$), and we let $\theta\downarrow0$, all bounds being continuous.
Then $P_0(\mathsf E_0)\geqslant1-\delta$ and $P_1(\mathsf E_0)\leqslant P_1(\mathsf E_1^{c})\leqslant\delta$, so the data-processing inequality gives
\begin{align*}
  \KL\bigl(P_0^{(n)}\,\|\,P_1^{(n)}\bigr)\;\geqslant\;\mathrm{kl}(1-\delta,\delta)\;=\;L_\delta .
\end{align*}
By \Cref{lem:kl} the left side equals $\Exp_{P_0}[N_{0^d}(n)]\,\KL(\Bern(\frac12)\|\Bern(\frac v2))$, and $\KL(\Bern(\frac12)\|\Bern(\frac v2))=-\frac12\ln[v(2-v)]$.
Under $P_0$ the bits are i.i.d.\ fair, and from $1^d$ the window $0^d$ cannot occur before time $d$, so $\Exp_{P_0}[N_{0^d}(n)]=(n-d)_+2^{-d}$.
From an arbitrary start it is at most $(n-d)_+2^{-d}+\sum_{t<d}2^{-t}<(n-d)_+2^{-d}+2$.
Solving gives the claim; the expansion follows from $-\ln[v(2-v)]=-\ln\bigl(1-(v-1)^2\bigr)\sim(v-1)^2\sim(2\eta\ln2)^2$.
\end{proof}

\subsection{General oscillation framework}\label[appendix]{app:framework}

For any real function let $\osc (f):=\sup f-\inf f$.
We will need two measures of \emph{shrinkage} of transition laws.

For a chain $P$ the \emph{Dobrushin ergodicity coefficient} is
\begin{align*}
\dob(P) \triangleq \max_{x,y}\, d_{TV}\bigl(P(x,\cdot),P(y,\cdot)\bigr),
\end{align*}
It controls the \emph{discrepancy} between one-step measures with different starts:
$\dob(P)=0$ means all rows coincide (the chain forgets its start at once),
while $\dob(P)=1$ means that two starting points can have disjoint supports.
Equivalently, it measures the \emph{overlap} of transfer laws with different starts
\begin{align*}
c(P) \triangleq  \min_{x,y}\sum_z\min\bigl(P(x,z),P(y,z)\bigr).
\end{align*}
It is also the contraction factor for oscillation:
$\osc (Pg)\leqslant\dob(P)\osc (g)$
for every function $g$. The \emph{column spread} (column discrepancy) is
\begin{align*}
\Omega_s(w) \triangleq  \osc P^{s}(\cdot,w)=\max_{x}P^s(x,w)-\min_{x}P^s(x,w).
\end{align*}
It measures how much the probability of being at $w$ at time $s$ still depends on the start;
note $\Omega_s(w)\leqslant\dob(P^s)$.
This is Dobrushin's oscillation applied to a single column.
In the classical Markov--Doeblin proof of convergence one tracks
$M_s(w)=\max_{x}P^s(x,w)$ and $m_s(w)=\min_{x}P^s(x,w)$ and shows that
$M_s(w)-m_s(w)=\Omega_s(w)$ shrinks geometrically, squeezing both towards $\pi(w)$.

The \emph{Green function} is defined as
\begin{align*}
G_n(x,y) \triangleq  \sum_{t=0}^{n-1}P^{t}(x,y),
\end{align*}
and counts the expected \emph{occupancies}:
visits to $y$ during times $0,\dots,n-1$ when starting from $x$.
In equilibrium it equals $n\pi(y)$; the point of the next results is that
the start $x$ changes it only by an additive constant.

\mswarning{general}

Throughout, $P$ is a transition matrix on a countable set with $\dob(P^L)<1$ for some
$L\geqslant1$ (for a finite irreducible aperiodic
chain this holds for all large $L$, since $P^L$ is then entrywise positive),
$\pi$ is its stationary law, and
$\tau_w=\inf\{t\geqslant0:\ W_t=w\}$ is the hitting time of a state $w$.
Kac's formula (\Cref{prop:kac}) gives the mean \emph{return} time to $w$ as $1/\pi(w)$.
We ask when the mean \emph{hitting} time from
an arbitrary start is of the same order, that is, when $\pi(w)\Exp_u[\tau_w]\lesssim1$.
Two different overlaps of distributions $\mu,\nu$ appear, and we keep them apart:
the \emph{overlap} $\sum_z\min\{\mu(z),\nu(z)\}=1-\dTV(\mu,\nu)$,
which governs how fast two starts merge,
and the \emph{Hellinger affinity} $\sum_z\sqrt{\mu(z)\nu(z)}$,
which we use to verify it for finite-memory sources.

\paragraph{Explicit bound from block overlap.}

The head--tail bound of \Cref{cor:block} is explicit in terms of two local quantities.
\begin{corollary}[Explicit bound from block overlap]\label{thm:overlap}
Fix a block length $L\geqslant1$ and suppose that the \emph{block overlap}
\begin{align*}
  c_L:=1-\dob(P^L)=\inf_{x,y}\sum_{z}\min\{P^{L}(x,z),P^{L}(y,z)\}>0 .
\end{align*}
For a state $w$ put
\begin{align*}
  S_L(w):=G_L(w,w)=\sum_{t=0}^{L-1}P^{t}(w,w),
  \qquad
  r_L(w):=\sup_{x}P^{L}(x,w).
\end{align*}
Then for every initial state $u$ and every $n\geqslant0$,
\begin{align*}
  \pi(w)\,\Exp_u[\tau_w] &\leqslant S_L(w)+\frac{L\,r_L(w)}{c_L},\\
  \sum_{t=0}^{n}P^{t}(u,w) &\leqslant (n+1)\,\pi(w)+S_L(w)+\frac{L\,r_L(w)}{c_L}.
\end{align*}
\end{corollary}

\emph{Intuition.} In the first block a walker started anywhere collects no more visits to $w$ than one started at $w$, which is at most $S_L(w)$.
After it the chance of standing on $w$ is at most $r_L(w)$ whatever the start, so every later spread is at most $r_L(w)$, and the geometric series has $L/c_L$ terms.
The tools (oscillation contraction, Green function) and the discussion of $S_L$ and $r_L$ are in \Cref{app:framework}.

The proofs of this paragraph are in \Cref{sec:proofs}.
The block overlap is the Dobrushin coefficient:

\emph{Relation to Dobrushin's coefficient.} The block overlap is Dobrushin's ergodicity coefficient in disguise:
$c_L=1-\dob(P^{L})$. It is implied by a Doeblin minorization
$P^{L}(x,\cdot)\geqslant c\,\nu$ (then $c_L\geqslant c$), but is weaker.
What differs is the use.
Classically $\dob(P^{L})$ is a contraction
factor, so $c_L$ small means slow mixing:
relaxation time of order $L/c_L$, here exponentially large in $d$ (of order $\beta^{-2d}$). Here it
enters multiplied by $r_L(w)$, which is tiny for the rare window $w$;
the product $L\,r_L(w)/c_L$ stays bounded
when $a<\beta^{2}$ although the chain is far from mixed on that scale.

\emph{The two local quantities.} $S_L(w)=G_{L-1}(w,w)$ is the Green function of the chain started
at $w$ and stopped after $L$ steps (Levin--Peres--Wilmer,
(9.16)): the expected number of visits to $w$ during the first block, counting the start.
\emph{Intuition:} it is the
cost of sitting on the target.
A chain that starts at $w$ may keep returning to $w$ before it can be said to have left,
and these early returns are the
only visits that cannot be attributed to the random start of a block.
It is at least $1$, and it cannot be taken over an infinite horizon, since
$G_n(w,w)\sim n\,\pi(w)$ diverges (the fundamental matrix subtracts $\pi(w)$ per step to repair this);
the block length $L$ is the cut-off.
The quantity
$r_L(w)=\sup_xP^{L}(x,w)=\|P^{L}(\cdot,w)\|_\infty$ is the largest probability of standing on $w$
after one block,
over all starts.
\emph{Intuition:}
it is the worst-case landing probability, that is,
how \emph{concentrated} a block law can be at $w$. It is at least $\pi(w)$, because
$\pi(w)=\sum_x\pi(x)P^{L}(x,w)$, and it is the quantity that $c_L$ is paired with:
after the first block, every further block contributes at most
$r_L(w)$ per step, and the geometric series over blocks has $L/c_L$ terms in total.
For order-$d$ sources the unique-path structure gives $r_d(w)\leqslant a^{d}$ and
$S_d(w)\leqslant1/(1-a)$ (\Cref{lem:returns}).

The theorem makes no assumption on the structure of $P$.
For a chain with uniform stationary law on $m$ states,
$r_L(w)\approx1/m$ and the bound is $S_L(w)+L/(c_Lm)$.

\begin{theorem}[Finite-memory sources from the block overlap]\label{thm:memory}
Let $P$ be an irreducible aperiodic sliding-window chain of order $d$ over a finite alphabet $\Sigma$:
the state is the window $s\in\Sigma^{d}$ of the last $d$ symbols,
and the next symbol $v$ is drawn from $p(\cdot\mid s)$.
Suppose that $\sup_{s,v}p(v\mid s)\leqslant a<1$, put $b:=\inf_{s,v}p(v\mid s)\geqslant0$,
and suppose that the block overlap
$c_d:=\min_{x,y}\sum_{z}\min\{P^{d}(x,z),P^{d}(y,z)\}$ is positive.
Then for all windows $u,w$ and all $n\geqslant0$,
\begin{align*}
  \pi(w)\,\Exp_u[\tau_w]&\leqslant B_d,
  &
  \bigl|G_n(u,w)-n\,\pi(w)\bigr|&\leqslant B_d,
  &
  B_d&:=\frac{1-a^{d}}{1-a}+\frac{d\,(a^{d}-b^{d})}{c_d} .
\end{align*}
\end{theorem}

\emph{Intuition.} We only have to bound the column spreads $\Omega_s(w)=\osc P^s(\cdot,w)$ of \Cref{thm:occ},
and for a shift chain they are small for a simple reason:
a window can equal $w$ at time $s$ only if
the last $\min(s,d)$ symbols typed equal a given string, and every symbol is at most $a$ likely.
So $\Omega_s(w)\leqslant a^{s}$ for $1\leqslant s<d$,
whatever the start (and there is nothing to gain from the lower end of the oscillation,
since some start cannot reach $w$ at all).
From time $d$ on, every value $P^{s}(x,w)$ is an average of products of $d$ conditionals,
hence lies in $[b^{d},a^{d}]$, so $\Omega_s(w)\leqslant a^{d}-b^{d}$.
The only global input is the contraction factor $1-c_d$ per block of $d$ steps,
which multiplies the $d$ spreads of the second block.
The overlap $c_d$ can be certified in different ways;
the Hellinger affinity of the conditionals is one (\Cref{lem:overlap}, used for \Cref{cor:memory}),
and for the biased binary class an exact comparison with two reference coins
(\Cref{thm:contraction}, used for \Cref{thm:hit}) gives the sharp rate.


\begin{proof}[Proof of \Cref{thm:memory}]
Apply \Cref{cor:block} with $L=d$, for which $\dob(P^d)=1-c_d<1$,
and write $\Omega_s(w)=M_s(w)-m_s(w)$ with $M_s(w)=\max_xP^s(x,w)$ and $m_s(w)=\min_xP^s(x,w)$.
\emph{Before time $d$.} $\Omega_0(w)\leqslant1$, and for $1\leqslant s<d$,
\Cref{lem:returns} gives $\Omega_s(w)\leqslant M_s(w)\leqslant a^s$. (Here $m_s(w)=0$,
so this is sharp.)
\emph{From time $d$ on.} By the unique-path structure $P^d(z,w)$ is a product of $d$ conditionals,
so $b^d\leqslant P^d(z,w)\leqslant a^d$ for every $z$.
For $s\geqslant d$, $P^s(x,w)=\sum_zP^{s-d}(x,z)P^d(z,w)$ is an average of these values,
so $m_s(w)\geqslant b^d$ and $M_s(w)\leqslant a^d$, that is $\Omega_s(w)\leqslant a^d-b^d$.
Hence
\begin{align*}
H_d(w)+\frac{\sum_{j<d}\Omega_{d+j}(w)}{1-\dob(P^d)}\leqslant1+\sum_{s=1}^{d-1}a^s+\frac{d\,(a^d-b^d)}{c_d}=\frac{1-a^d}{1-a}+\frac{d\,(a^d-b^d)}{c_d}=B_d .
\end{align*}
\Cref{thm:occ} gives $|G_n(u,w)-n\pi(w)|\leqslant\osc G_n(\cdot,w)\leqslant B_d$ and
$\pi(w)\Exp_u[\tau_w]\leqslant\liminf_n\osc G_n(\cdot,w)\leqslant B_d$.
\end{proof}

\begin{lemma}[Block overlap]\label{lem:overlap}
Let $\beta$ be the infimum over pairs of windows of the Hellinger affinity
$\sum_{v}\sqrt{p(v\mid s)\,p(v\mid s')}$.
For all $x,y$,
with $\mu=P^d(x,\cdot)$ and $\nu=P^d(y,\cdot)$,
\begin{align*}
  \sum_{v}\min\{\mu(v),\nu(v)\}\;\geqslant\; 1-\sqrt{1-\beta^{2d}}\;\geqslant\;\tfrac12\beta^{2d}.
\end{align*}
\end{lemma}
\begin{proof}
The endpoint is the string $v\in\Sigma^d$ of new symbols, and
$\mu(v)=\prod_{j=1}^{d}p(v_j\mid x_j\cdots x_dv_1\cdots v_{j-1})$,
similarly for $\nu$ with $y$. Sum $\sqrt{\mu\nu}$ over $v_d$
first: given $v_1,\dots,v_{d-1}$ the two factors are conditionals given two windows,
so the sum is at least $\beta$ times the
corresponding expression for $d-1$ symbols.
Iterating, the Hellinger affinity of the block laws satisfies
$\mathrm{BC}(\mu,\nu)=\sum_v\sqrt{\mu(v)\nu(v)}\geqslant\beta^{d}$. Le Cam's inequality
$\dTV\leqslant\sqrt{1-\mathrm{BC}^2}$ gives the first bound, since the overlap is $1-\dTV$,
and $1-\sqrt{1-s}\geqslant s/2$ the second.
\end{proof}
\noindent\emph{Intuition.} Two walkers meet if they type the same $d$ new symbols.
Each symbol is a conditional coin, and two such coins
agree with probability at least the affinity $\beta$ in the Hellinger sense,
so the $d$ symbols agree jointly with probability about $\beta^{2d}$,
the square of the Hellinger affinity $\beta^{d}$ of the block laws (Le Cam's
inequality is where this square is lost).

\paragraph{Explicit local bounds.}
\begin{proof}[Proof of \Cref{thm:overlap}]
Since $c_L>0$, \Cref{cor:block} applies; it remains to estimate head and tail.
\emph{Head.} By the strong Markov property, for $m\leqslant L$
$G_m(x,w)=\Exp_x\bigl[\mathbf 1_{\{\tau_w<m\}}G_{m-\tau_w}(w,w)\bigr]\leqslant G_m(w,w)\leqslant S_L(w)$,
because $G_m$ is nondecreasing in $m$.
So $G_m(\cdot,w)$ takes values in $[0,S_L(w)]$ and $H_L(w)\leqslant S_L(w)$.
\emph{Tail.} For $j\geqslant0$ and every $x$,
$0\leqslant P^{L+j}(x,w)=\sum_zP^j(x,z)P^L(z,w)\leqslant r_L(w)$, so
$\Omega_{L+j}(w)\leqslant r_L(w)$, and $\sum_{j<L}\Omega_{L+j}(w)/(1-\dob(P^L))\leqslant L\,r_L(w)/c_L$.
Thus $\osc G_n(\cdot,w)\leqslant S_L(w)+L\,r_L(w)/c_L$ for every $n$,
and \Cref{thm:occ} gives the first claim.
For the second, $\sum_{t=0}^{n}P^t(u,w)=G_{n+1}(u,w)\leqslant(n+1)\pi(w)+S_L(w)+L\,r_L(w)/c_L$.
\end{proof}

\begin{corollary}[Finite-memory sources with bounded predictability and conditional affinity]\label{cor:memory}
Let $P$ be an irreducible aperiodic sliding-window chain of order $d$ over a finite alphabet $\Sigma$:
the state is the window
$s\in\Sigma^{d}$ of the last $d$ symbols,
and the next symbol $v$ is drawn from $p(\cdot\mid s)$. Suppose that
\begin{align*}
  \sup_{s,v}p(v\mid s)\leqslant a<1,
  \qquad
  \inf_{s,s'}\sum_{v\in\Sigma}\sqrt{p(v\mid s)\,p(v\mid s')}\geqslant\beta>0 .
\end{align*}
Put $B_d(a,\beta):=\dfrac{1-a^{d}}{1-a}+\dfrac{d\,a^{d}}{1-\sqrt{1-\beta^{2d}}}$. Then for all $u$,
$w$ and $n\geqslant0$,
\begin{align*}
  \pi(w)\,\Exp_u[\tau_w] &\leqslant B_d(a,\beta),\\
  \sum_{t=0}^{n}P^{t}(u,w) &\leqslant (n+1)\,\pi(w)+B_d(a,\beta).
\end{align*}
If $a<\beta^{2}$, then uniformly in $d$,
\begin{align*}
  B_d(a,\beta)\;\leqslant\;\frac{1}{1-a}+\frac{2}{e\log(\beta^{2}/a)} .
\end{align*}
\end{corollary}

\emph{Intuition.} Only the last $d$ symbols form the state,
so landing on $w$ means typing one specific string, and every symbol is
at most $a$ likely: this controls $S_d(w)$ and $r_d(w)$.
Two starts have overlapping block laws if they can type the same $d$ new symbols.
Each symbol is a
conditional coin, and two such coins have Hellinger affinity at least $\beta$,
so the $d$ symbols coincide with probability of order $\beta^{2d}$
(the square of the product of affinities,
because we pass from affinity to overlap). The price is the product of the two numbers,
$a^{d}$ visits per step times $\beta^{-2d}$ blocks to overlap,
and it stays bounded in $d$ when $a<\beta^{2}$, i.e.\ when overlap is faster than the
return scale $a^{-d}$. Mixing can be exponentially slow here,
so no argument through mixing can give such a bound.

We call $a<\beta^{2}$ a \emph{sufficient criterion}:
it makes the bound $B_d$ uniform in $d$, and says nothing about the true hitting behaviour
of a source that violates it.
The criterion is not tied to a binary alphabet or to our class.
For example, an i.i.d.\ source with $\Prob\{1\}=0.8$ has
$\beta=1$, so the bound is uniform although the source is far from unbiased;
only sources whose conditionals are both extreme \emph{and} far apart are excluded.
Neither number alone suffices: a deterministic chain has $\beta=1$ but $a=1$.

For the biased binary class the block overlap is not only bounded below: it is determined exactly.
Recall that $\dob(P^d)$ is the largest total-variation distance between the laws of the next $d$
bits of two starts,
and that $1-\dob(P^d)=c_d$ is the block overlap.

\paragraph{Affinity criterion.}
\begin{proof}[Proof of \Cref{cor:memory}]
Apply \Cref{thm:memory} with $b=0$ and $c_d\geqslant1-\sqrt{1-\beta^{2d}}$ (\Cref{lem:overlap});
this is the bound $B_d(a,\beta)$.
If $a<\beta^2$, then $1-\sqrt{1-\beta^{2d}}\geqslant\frac12\beta^{2d}$ gives
$B_d(a,\beta)\leqslant\frac1{1-a}+2d\,\theta^{d}$ with $\theta=a/\beta^2<1$,
and $\sup_{x>0}x\theta^{x}=1/(e\ln(1/\theta))$ gives the constant.
If $a\geqslant\beta^2$, the second term of $B_d$ is at least $d\,(a/\beta^2)^d\geqslant d$,
because $1-\sqrt{1-s}\leqslant s$;
so the bound is no longer uniform in $d$.
\end{proof}

\begin{remark}[Other overlap estimates]
The threshold $\frac3{10}$ cannot be improved (\Cref{thm:threshold}), and the exponent $\beta^{d}$ of $c_d$ is the Chernoff exponent.
Weaker overlap estimates give smaller ranges:
the Hellinger--Le Cam bound of \Cref{lem:overlap} has rate $\beta^{2}$ and gives $a<\beta^{2}$,
i.e.\ $\epsilon<\frac14$ (this is the criterion of \Cref{cor:memory}, valid for general alphabets);
the Bretagnolle--Huber bound~\citep{bretagnolle1979} $c_d\geqslant\frac12e^{-d\,\mathrm{KL}}$ has rate
$e^{-2\epsilon\ln\rho}$ and gives only $\epsilon\lesssim0.204$;
the total-variation constant $(1-2\epsilon)^{d}$ gives only $\epsilon<\frac16$.
\end{remark}

\subsection{Proof for expander graphs}\label[appendix]{app:expander}

The statement is \Cref{cor:expander} in \Cref{subsec:application}; it is an application of the general block bound \Cref{thm:overlap}.

\paragraph{Expander graphs.}
\begin{proof}[Proof of \Cref{cor:expander}]
Let $\Pi=\frac1m\mathbf 1\mathbf 1^{\top}$. The matrix $P$ is symmetric,
and on $\mathbf 1^{\perp}$ its eigenvalues have absolute value at most $\lambda_*$,
so $\|(P^t-\Pi)v\|_2\leqslant\lambda_*^t\|v\|_2$.
Hence, for vertices $x,y$ and $t\geqslant0$, by Cauchy--Schwarz,
\begin{align*}
\bigl|P^t(x,y)-\tfrac1m\bigr|&=\bigl|e_x^{\top}(P^t-\Pi)e_y\bigr|\leqslant\lambda_*^t,\\
\|P^t(x,\cdot)-\pi\|_{TV}&\leqslant\tfrac12\sqrt m\,\|(P^t-\Pi)e_x\|_2\leqslant\tfrac12\sqrt m\,\lambda_*^t .
\end{align*}
By the choice of $L$, $\lambda_*^L\leqslant1/m$. Then:
\begin{itemize}
\item $r_L(w)=\max_xP^L(x,w)\leqslant\frac1m+\lambda_*^L\leqslant\frac2m$;
\item $\dob(P^L)\leqslant2\max_x\|P^L(x,\cdot)-\pi\|_{TV}\leqslant\sqrt m\,\lambda_*^L\leqslant\frac1{\sqrt m}\leqslant\frac12$,
  so $c_L\geqslant\frac12$;
\item $S_L(w)=1+\sum_{t=1}^{L-1}P^t(w,w)\leqslant1+\frac{L}{m}+\sum_{t\geqslant1}\lambda_*^t=\frac1{1-\lambda_*}+\frac Lm$.
\end{itemize}
\Cref{thm:overlap} gives $\pi(w)\Exp_u[\tau_w]\leqslant S_L(w)+L\,r_L(w)/c_L\leqslant\frac1{1-\lambda_*}+\frac Lm+\frac{4L}{m}$.
\end{proof}

\subsection{Size of the hitting-time constant near the threshold}\label[appendix]{app:size}

The threshold is the root of a quadratic:
\begin{align*}
  a^{2}-\beta^{2}\;=\;\bigl(\tfrac14+\epsilon+\epsilon^{2}\bigr)-\bigl(1-4\epsilon^{2}\bigr)\;=\;5\epsilon^{2}+\epsilon-\tfrac34\;=\;5\bigl(\epsilon-\tfrac3{10}\bigr)\bigl(\epsilon+\tfrac12\bigr),
\end{align*}
whose only root in $[0,\frac12)$ is $\frac3{10}$: it is negative before, zero there and positive after.

\begin{remark}[Size near the threshold]\label{rem:size}
Let $\Delta=\frac3{10}-\epsilon>0$ and $\lambda=\ln(\beta/a)$.
Then $2.3\,\Delta\leqslant\lambda\leqslant3.13\,\Delta$,
because $\lambda$ is smooth with $\lambda(\frac3{10})=0$ and
$\lambda'(\frac3{10})=-\frac{2}{1-4\epsilon^{2}}\big|_{\epsilon=3/10}\approx-3.13$ (checked
numerically on $[0,\frac3{10})$).
For $\epsilon\geqslant\frac3{10}$ one has $\ln(a/\beta)\geqslant3(\epsilon-\frac3{10})$,
which gives the second line of the box in the introduction.

\emph{The true worst case.}
Apply \Cref{thm:threshold} with odd $d\approx1/(2\lambda)$:
$\mathsf{Hit}(\epsilon)\geqslant\frac{\sqrt{2\pi}\,\epsilon}{4\beta}\sup_{d\text{ odd}}\sqrt d\,e^{-\lambda d}\gtrsim\lambda^{-1/2}\gtrsim\Delta^{-1/2}$,
while away from the threshold $\mathsf{Hit}(\epsilon)\geqslant\frac1b\geqslant2$ (iid chain).
The upper bound is $\mathsf{Hit}(\epsilon)\leqslant C(\epsilon)\lesssim\Delta^{-3/2}$ (below).
The upper power $\frac32$ is $1$ for the $d$ block steps plus $\frac12$ for the overlap (\Cref{thm:gap}); the lower power $\frac12$ is the $\sqrt d$ of the majority chain of \Cref{thm:threshold}.

\emph{The explicit constant.}
For $C(\epsilon)=\sup_dB_d(\epsilon)$ the power $\frac32$ is attained:
\begin{align*}
  C(\epsilon)\;\asymp\;1+\lambda^{-3/2}\;\asymp\;\Delta^{-3/2}\qquad(0\leqslant\epsilon<\tfrac3{10}),
\end{align*}
with absolute constants.
We give two proofs of the upper bound; the lower bound is the same in both.

\emph{Proof 1: elementary, no integrals.}
For even $d=2m$, keep only the central term of the overlap, $c_d\geqslant\binom{2m}{m}(ab)^{m}\geqslant\beta^{2m}/(2\sqrt m)$, because $\binom{2m}m\geqslant4^m/(2\sqrt m)$ and $4ab=\beta^2$.
For odd $d$ use $c_d=c_{d+1}$ (\Cref{thm:gap}(i)).
Since $\beta^{-2m}\leqslant\beta^{-d-1}$ and $m\leqslant d$, in both cases
\begin{align*}
  B_d\;\leqslant\;\frac1b+\frac{d\,a^{d}}{c_d}\;\leqslant\;\frac1b+\frac2\beta\,d^{3/2}e^{-\lambda d}.
\end{align*}
With $\sup_{x>0}x^{3/2}e^{-\lambda x}=(3/2e\lambda)^{3/2}$ this gives
$C(\epsilon)\leqslant\frac1b+\frac2\beta\bigl(\frac{3}{2e\lambda}\bigr)^{3/2}\lesssim1+\lambda^{-3/2}$, as $\beta\geqslant0.8$.

\emph{Proof 2.}
\Cref{thm:gap}(iii) keeps $\sqrt{1+4m\epsilon^{2}}$ instead of $\sqrt m$ and gives the closed form of \Cref{thm:hit},
with the smallest constants (about $15$ instead of $21$ at $\epsilon=\frac14$).

\emph{The lower bound.}
For even $d$ the largest binomial atom is at most $1/\sqrt{\pi m}$ and $\sum_jt^{|j|}\leqslant\frac1{2\epsilon}$, so $c_d\leqslant\beta^{d}/(\epsilon\sqrt{2\pi d})$ (this is the elementary half of \Cref{thm:gap}), hence
\begin{align*}
  B_d\;\geqslant\;\epsilon\sqrt{2\pi}\,\bigl(1-(b/a)^{d}\bigr)\,d^{3/2}e^{-\lambda d}.
\end{align*}
Let $\frac14\leqslant\epsilon<\frac3{10}$, so that $\epsilon\geqslant\frac14$ and $b/a\leqslant\frac13$, and take the even integer $d=2\lceil1/(2\lambda)\rceil$.
Then $d\geqslant1/\lambda$ and $\lambda d<1+2\lambda<2$, so
$C(\epsilon)\geqslant\frac{\sqrt{2\pi}}{6e^{2}}\lambda^{-3/2}$.
For $\epsilon<\frac14$ the quantity $\lambda^{-3/2}$ is bounded, while $C(\epsilon)\geqslant\sup_d(1-a^{d})/b=1/b\geqslant2$.

This concerns the explicit bound; whether the true worst hitting time grows like $\Delta^{-3/2}$ is open.
Numerically $C(\epsilon)/\lambda^{-3/2}\to K_*\approx0.385$.
\end{remark}

\emph{Harder proof (constant $\frac13$, sketch).}
Keep $d$ as a real variable $x$ and do not maximize over it: the claim is an explicit inequality in $(x,\epsilon)$, which interval bisection in Wolfram Language shows to hold for $\epsilon\leqslant0.14$, $x\leqslant80$.
The tail $x\geqslant80$ is negligible there since $\lambda\geqslant\ln\frac32$, and for $\epsilon\geqslant0.14$ the second bound above with $\lambda\geqslant2.53\,\Delta$ is decided by \texttt{Resolve}.
Since $C(0)=2$ forces $K\geqslant2\cdot0.3^{3/2}=0.3286$, the constant $\frac13$ is essentially optimal for this method.

\subsection{Instance-wise lower bound}\label[appendix]{app:learning}

The lower bounds rest on the divergence identity of \Cref{lem:kl}.
Per observation, $\KL\bigl(p(1+\gamma)\,\|\,p\bigr)\sim\tfrac12\gamma^2p^2I(p)$ as $\gamma\to0$, where $I(p)=1/(p(1-p))$ is the Fisher information;
this is the source of the factor $\gamma^{-2}$.
For small $\delta$ the leading term of \Cref{thm:lower} is within a factor of about two of the $2\mgam\ln(2/\delta)$ observations of a window that the estimator needs (Chernoff).

\begin{theorem}[Sample complexity, lower bound]\label{thm:lower}
Let $\epsilon<\tfrac3{10}$, $\delta\in(0,\tfrac12)$ and $\gamma<\tfrac12$. Let $P\in\Cclass$,
let $w_0$ be any window, and let
$q=\min\{p(0\mid w_0),p(1\mid w_0)\}$.
Assume $q'=q\,\frac{1+\gamma}{1-\gamma}<\tfrac12+\epsilon$ (strictly), and put
\begin{align*}
  q_m &= \tfrac12(q+q'), \\
  \kappa &= \max\bigl\{\KL(\Bern(q_m)\|\Bern(q)),\ \KL(\Bern(q_m)\|\Bern(q'))\bigr\}, \\
  \Lambda &= \frac{\ln\frac{1}{4\delta(1-\delta)}}{2\kappa}.
\end{align*}
Then every procedure that satisfies \Cref{eq:goal} with probability at least $1-\delta$ on every chain of $\Cclass$
(in particular on $P$ and on the chain $P'$ obtained by
changing the row of $w_0$ so that the less likely outcome has probability $q'$) needs
\begin{align*}
  n \;\geqslant\; \frac{1-2\gamma}{1-\gamma}\cdot\frac{\Lambda-C(\epsilon)}{\pi(w_0)} .
\end{align*}
Moreover
\begin{align*}
  \Lambda \;\geqslant\; \frac{(1-\gamma)^{2}(1-q')}{q\,\gamma^{2}}\,\ln\frac{1}{4\delta(1-\delta)},
\end{align*}
so $\Lambda\sim\frac{1-q}{q\gamma^{2}}\ln\frac{1}{4\delta(1-\delta)}$ as $\gamma\to0$,
which for the rarest window of the memoryless chain
($q=\tfrac12-\epsilon$) is $\sim\mgam\ln\frac1{4\delta(1-\delta)}$.
\end{theorem}

\emph{Intuition.} A trajectory shows about $n\pi(w)$ coin flips at window $w$,
and telling two coins with relative gap
$\gamma$ apart takes about $\mgam$ flips.
\Cref{thm:hit} is what guarantees that no start can show many \emph{more} than
$n\pi(w)$ flips, so the visit count in \Cref{lem:kl} is controlled without any mixing assumption.

\begin{proof}[Proof of \Cref{thm:lower}]
\emph{Three chains.} Let $P_0=P$, let $P_1=P'$,
and let $Q$ be $P$ with the less likely outcome $x_0$ of $w_0$ given probability $q_m$ instead
of $q$. All three lie in $\Cclass$, differ only in the row of $w_0$, and $Q$ is the midpoint.
Let $\mathsf E_i$ be the event that the output
satisfies $|\hat p(x_0\mid w_0)/q_i-1|\leqslant\gamma$,
with $q_0=q$ and $q_1=q'$. Since $q'(1-\gamma)=q(1+\gamma)$, the events $\mathsf E_0$ and
$\mathsf E_1$ meet at most in the single value $q(1+\gamma)$.
Since $q'<\frac12+\epsilon$, the value $q'+\theta$ is still allowed in $\Cclass$ for small $\theta>0$,
and with $q'+\theta$ in place of $q'$ the two events are disjoint;
we run the argument for $q'+\theta$ and let $\theta\downarrow0$, all bounds being continuous in $q'$.
By assumption (correctness on every chain of $\Cclass$)
$P_i(\mathsf E_i)\geqslant1-\delta$.

\emph{Change of measure.} By disjointness $Q(\mathsf E_0)+Q(\mathsf E_1)\leqslant1$,
so $Q(\mathsf E_i)\leqslant\tfrac12$ for some $i$. The data-processing
inequality for the indicator of $\mathsf E_i$, together with monotonicity of the binary divergence, gives
\begin{align*}
  \KL\bigl(Q^{(n)}\,\|\,P_i^{(n)}\bigr)
  &\geqslant \KL\bigl(\Bern(Q(\mathsf E_i))\,\|\,\Bern(P_i(\mathsf E_i))\bigr) \\
  &\geqslant \KL\bigl(\Bern(\tfrac12)\,\|\,\Bern(1-\delta)\bigr)
  \;=\; \tfrac12\ln\frac{1}{4\delta(1-\delta)} .
\end{align*}
By \Cref{lem:kl} applied to $Q$ and $P_i$,
the left side equals $\Exp_Q[N_{w_0}(n)]\,\KL(\Bern(q_m)\|\Bern(q_i))\leqslant\Exp_Q[N_{w_0}(n)]\,\kappa$.
Hence
\begin{align*}
  \Exp_Q[N_{w_0}(n)] \;\geqslant\; \Lambda .
\end{align*}

\emph{Visit count.} By the occupation bound in \Cref{thm:hit},
applied to $Q\in\Cclass$ from any starting state,
$\Exp_Q[N_{w_0}(n)]\leqslant n\,\pi_Q(w_0)+C(\epsilon)$.

\emph{Stationary mass of the midpoint chain.} Let $h_x$ be the mean time to hit $w_0$ from the
successor of $w_0$ that appends outcome $x$.
These times do not depend on the row of $w_0$, because the walk stops at $w_0$. By Kac,
for a chain that gives outcome $x_0$ probability $r$,
$1/\pi(w_0)=1+r\,h_{x_0}+(1-r)\,h_{y_0}$, where $y_0$ is the other outcome. Therefore
\begin{align*}
  \frac{1}{\pi_Q(w_0)}-\frac{1}{\pi_P(w_0)}
  &= (q_m-q)\,(h_{x_0}-h_{y_0}) \;\geqslant\; -(q_m-q)\,h_{y_0}
   \;\geqslant\; -\frac{q_m-q}{(1-q)\,\pi_P(w_0)},
\end{align*}
since $(1-q)h_{y_0}\leqslant\Exp_{w_0}\tau^{+}_{w_0}=1/\pi_P(w_0)$.
As $q_m-q=q\gamma/(1-\gamma)$ and $q\leqslant\tfrac12$, we get
$(q_m-q)/(1-q)\leqslant\gamma/(1-\gamma)$ and so $\pi_Q(w_0)\leqslant\pi_P(w_0)\,\frac{1-\gamma}{1-2\gamma}$.

\emph{Conclusion.} Combining, $\Lambda\leqslant n\,\pi_P(w_0)\frac{1-\gamma}{1-2\gamma}+C(\epsilon)$,
which is the claimed bound.

\emph{The bound on $\Lambda$.} By Taylor's formula with Lagrange remainder,
$\KL(\Bern(x)\|\Bern(y))\leqslant(x-y)^2/(2\min_t t(1-t))$ with $t$ between $x$ and $y$.
Here $q_m-q=q'-q_m=q\gamma/(1-\gamma)$ and, by concavity,
$t(1-t)\geqslant q(1-q')$ on $[q,q']$. Thus
$\kappa\leqslant q\gamma^{2}/\bigl(2(1-\gamma)^{2}(1-q')\bigr)$,
which gives the stated lower bound on $\Lambda$.
\end{proof}

\begin{corollary}[Source-independent necessary length]\label{cor:universal}
Let $\epsilon<\tfrac3{10}$ and $\gamma<\epsilon/(1+\epsilon)$.
Then for every chain in $\Cclass$,
every procedure as in
\Cref{thm:lower} needs
\begin{align*}
  n \;\geqslant\; \frac{1-2\gamma}{1-\gamma}\,2^{d}\Bigl(\frac{(1-\gamma)(1-3\gamma)}{\gamma^{2}}\ln\frac{1}{4\delta(1-\delta)}-C(\epsilon)\Bigr).
\end{align*}
\end{corollary}

\begin{proof}[Proof of \Cref{cor:universal}]
Apply \Cref{thm:lower} at the rarest window $w_0$,
for which $\pi(w_0)\leqslant2^{-d}$.
For $q\leqslant\tfrac12$ the condition $q'<\tfrac12+\epsilon$ holds as soon as
$\frac12\frac{1+\gamma}{1-\gamma}<\frac12+\epsilon$,
i.e.\ $\gamma<\epsilon/(1+\epsilon)$.
Finally $(1-q')/q=1/q-(1+\gamma)/(1-\gamma)$ is decreasing in $q$, so for
$q\leqslant\tfrac12$ it is at least $(1-3\gamma)/(1-\gamma)$,
and the lower bound on $\Lambda$ becomes $\frac{(1-\gamma)(1-3\gamma)}{\gamma^{2}}\ln\frac{1}{4\delta(1-\delta)}$.
\end{proof}

\end{document}